\documentclass[aps,a4paper,superscriptaddress,nofootinbib,twocolumn,journal,twoside]{revtex4-1}
\usepackage{verbatim}   % notes
\usepackage{graphicx}% Include figure files
\usepackage{subfigure}
\usepackage{dcolumn}% Align table columns on decimal point
\usepackage{bm}% bold math
\usepackage{booktabs}
\usepackage{makecell}
\usepackage{hyperref}

\usepackage[table]{xcolor}
\usepackage{bm}
\usepackage{amsmath,amsfonts,amssymb}
\usepackage{acronym}
\usepackage{array}
\usepackage{multirow}
\usepackage{soul}
\usepackage{comment}
\usepackage{tabularx}
\usepackage{verbatim}
\usepackage{tikz}
\usepackage{rotating}
\usetikzlibrary{shapes.geometric, arrows}

\tikzstyle{startstop} = [rectangle, rounded corners, minimum width=3cm, minimum height=0.5cm,text centered, draw=black, fill=red!30]
\tikzstyle{process} = [rectangle, minimum width=3em, minimum height=1em, text centered, draw=black, fill=orange!30]
\usepackage{ulem}

\newcommand{\TRC}{
MOE Key Laboratory of TianQin Mission, TianQin Research Center for Gravitational Physics \& School of Physics and Astronomy, Frontiers Science Center for TianQin, Gravitational Wave Research Center of CNSA, Sun Yat-sen University (Zhuhai Campus), Zhuhai 519082, China
}
\newcommand{\Hebeiuni}{
Department of Physics, Hebei University, Baoding, 071002, China}
\newcommand{\baoding}{
Hebei Key Laboratory of High-precision Computation and Application of Quantum Field Theory, Baoding, 071002, China}

\newcommand{\phyhe}{Hebei Research Center of the Basic Discipline for Computational Physics, Baoding, 071002, China}

\definecolor{cerulean}{rgb}{0.0, 0.48, 0.65}

\begin{document}

%\preprint{APS/123-QED}

\title{Identification of gravitational waves from extreme mass ratio inspirals}% Force line breaks with \\

\author{Chang-Qing Ye}
\affiliation{\TRC}

\author{Hui-Min Fan}
\email{fanhm3@mail.sysu.edu.cn}
\affiliation{\Hebeiuni}
\affiliation{\baoding}
\affiliation{\phyhe}

 \author{Alejandro Torres-Orjuela}
\affiliation{\TRC}

\author{Jian-dong Zhang}
\affiliation{\TRC}

\author{Yi-Ming Hu}
\email{huyiming@mail.sysu.edu.cn}
\affiliation{\TRC}

\date{\today}% It is always \today, today,

%%%%%%%%%%%%%%%%%%%%%%%%%%%%%%%%%%%%%%%%%%%%%%%%%%%%%%%%%%%%%%%%%%%%%%%%%%%%%%%%%%%%%%%%%%%%%%

\begin{abstract}

% \Yecq{what problem did we solve.}
% \Yecq{what the reslut?}

Space-based gravitational wave detectors like TianQin or LISA could observe extreme-mass-ratio-inspirals (EMRIs) at millihertz frequencies. The accurate identification of these EMRI signals from the data plays a crucial role in enabling in-depth study of astronomy and physics.
We aim at the identification stage of the data analysis, with the aim to extract key features of the signal from the data, such as the evolution of the orbital frequency, as well as to pinpoint the parameter range that can fit the data well for the subsequent parameter inference stage. 
In this manuscript, we demonstrate the identification of EMRI signals without any additional prior information on physical parameters. 
High-precision measurements of EMRI signals have been achieved, using a hierarchical search.
It combines the search for physical parameters that guide the subsequent parameter inference, and a semicoherent search with phenomenological waveforms that reaches precision levels down to $10^{-4}$ for the phenomenological waveform parameters $\omega_{0}$, $\dot{\omega}_{0}$, and $\ddot{\omega}_{0}$. As a result, we obtain measurement relative errors of less than 4\,\% for the mass of the massive black hole, while keeping the relative errors of the other parameters within as small as 0.5\,\%.

%%%%%%%%%%%%%%%%%%%%%%%%%%%%%%%%%%%%%%%%%%%%%%%%%%%%%%%%%%%%%%%%%%%%%%%%%%%%%%%%%%%%%%%%%%%%
\acrodef{MBH}[MBH]{Massive Black Hole}
\acrodef{sBH}{stellar-mass Black Hole}
\acrodef{MBBH}[MBBH]{Massive Binary Black Hole}
\acrodef{sBBH}{stellar-mass Binary Black Hole}
\acrodef{CO}[CO]{Compact Object}
\acrodef{DWD}[DWD]{Double White Dwarf}
\acrodef{BBH}[BBH]{Binary Black Hole}
\acrodef{BNS}[BNS]{Binary Neutron Star}
\acrodef{BH}[BH]{Black Hole}
\acrodef{NS}[NS]{Neutron Star}
\acrodef{WD}[WD]{White Dwarf}
\acrodef{GW}[GW]{Gravitational Wave}
\acrodef{EMRI}[EMRI]{Extreme Mass Ratio Inspiral}
\acrodef{CNN}{convolutional neural network}
\acrodef{SNR}[SNR]{Signal-to-Noise Ratio}
\acrodef{AGN}{Active Galactic Nuclei}
\acrodef{TDI}{Time Delay Interferometry}

\acrodef{AK}[AK]{Analytic Kludge}
\acrodef{AAK}[AAK]{Augmented Analytic Kludge}
\acrodef{NK}[NK]{Numerical Kludge}

\acrodef{CNN}{Convolutional Neural Network}
\acrodef{FAP}{False Alarm Probability}
\acrodef{TAP}{True Alarm Probability}
\acrodef{ROC}{Receiver Operator Characteristics}

\acrodef{ML}{Machine Learning}
\acrodef{ANN}{Artificial Neural Network}
\acrodef{NN}{Neural Network}
\acrodef{GAN}{Generative Adversarial Networks}
\acrodef{MMA}{Multi-Messenger Astronomy}

%%%%%%%%%%%%%%%%%%%%%%%%%%%%%%%%%%%%%%%%%%%%%%%%%%%%%%%%%%%%%%%%%%%%%%%%%%%%%%%%%%%%%%%%%%%%

\begin{comment}
\begin{description}
\item[Usage]
Secondary publications and information retrieval purposes.
\item[PACS numbers]
May be entered using the \verb+\pacs{#1}+ command.
\item[Structure]
You may use the \texttt{description} environment to structure your abstract;
use the optional argument of the \verb+\item+ command to give the category of each item. 
\end{description}
\end{comment}
\end{abstract}
%                              % Classification Scheme.
% \keywords{EMRI, gravitational-wave, detection, convolutional neural network, CNN }%Use showkeys class option if keyword
\maketitle
%\tableofcontents

\section{Introduction}

%%% Revised on September 18 (start)
%%%--------------------------------------------------------------------------------%%%
The successful detection of gravitational waves (GWs) has provided a new way to understand the Universe. To date, ground-based gravitational wave observatories such as LIGO and Virgo have detected nearly one hundred GW signals originating from the coalescence of compact binary systems at high frequencies ($10-1000$ Hz)~\cite{LIGOScientific:2018mvr,abbott2021gwtc,abbott2023gwtc}. Future space-based detectors, including TianQin and LISA, with longer armlengths  will detect heavier sources, such as ones involving massive black holes (MBHs), or even the low-frequency (subhertz) inspiral phase of stellar-origin compact binaries~\cite{danzmann2003lisa,AmaroSeoane:2012km,amaro2017laser,mei2020tianqin,torres-orjuela_huang_2023}.

% TianQin is a proposed space-borne gravitational-wave observatory consisting of three spacecraft in Earth orbit with arm lengths of about $1.7\times 10^{5} \rm km$,will be sensitive to GWs in the
% frequency band $10^{-4}-0.1 \rm Hz$\cite{luo2016tianqin,hu2018fundamentals,liu2020science}.

% \ato{I see no need to mention EM observation here but the state of GW detection could be extended.}

A prominent source that space-based detectors will detect is extreme mass ratio inspirals (EMRIs). These sources are formed by a stellar-mass compact object such as a stellar origin black hole (SOBH) with a mass $\mu \sim 1-10^2 \rm{M_{\odot}}$ inspiralling into a massive black hole (MBH) with a mass $M \sim 10^4-10^7 \rm{M_{\odot}}$\cite{AmaroSeoane:2007aw,AmaroSeoane:2012tx}. During the inspiral, EMRIs complete $\sim 10^4-10^5$ orbits and emit GWs that will allow us to study the space-time around the MBH with TianQin or LISA. Modeling and extracting EMRI signals from data streams will allow us to provide an intrinsic parameter estimation accuracy of $10^{-5}$ or even higher~\cite{babak2017science,gair2017prospects,fan2020science,Zi:2021pdp,torres-orjuela_huang_2023}. This precision will enable precise tests of general relativity\cite{Maselli:2021men,Barsanti:2022ana,torres2021exciting}, better understanding of the properties of MBHs~\cite{barausse_rezzolla_2008,Pan:2021oob,Fan:2022wio} and their surrounding environments\cite{Yunes2011,Barausse2015,cardoso2022gravitational}. Furthermore, gravitational
wave signals from EMRIs could be used to understand the mass
function of MBHs~\cite{Gair2010} and to constrain cosmological parameters~\cite{MacLeod:2007jd,Laghi2021EMRICosmology}.

Performing an end-to-end data analysis process for EMRIs is still a great challenge. This is mainly constrained by two aspects. First, it requires waveform templates that can be generated both quickly and highly accurately. Based on the Teukolsky equation and gravitational self-force calculations, one can obtain accurate EMRI waveforms but generating them is very expensive~\cite{hughes2005gravitational,self-force_Van,PT_Barack}. Therefore, data analysis often chooses computationally affordable alternatives such as kludge waveform~\cite{AK_2004_PRD,NK_2007_PRD,chua2017augmented} that are fast but inaccurate. For our analysis, we use FastEMRIWaveforms (FEW) in the Schwarzschild eccentric condition which is a state-of-the-art model that can rapidly generate fully relativistic EMRI waveforms while being more accurate than the aforementioned kludge waveforms~\cite{chua2021rapid,katz2021fast}. Second, data analysis methods pose significant challenges to achieving high-precision measurements for EMRIs. The multipeak structure of the posterior distribution in intrinsic parameter space makes it difficult for stochastic sampling algorithms to efficiently explore the parameter space, especially to enable transition between peaks~\cite{chua2022nonlocal}.
Meanwhile, the order-of-magnitude estimate shows that naively placing grids over the parameter space might require $\sim10^{40}$ templates~\cite{gair2004event}.

% On the other hand,the theoretical accuracy of parameter estimation is very high. TianQin or LISA can achieve a precision of $10^{-5}$ for intrinsic parameters\cite{babak2017science,gair2017prospects,fan2020science,Zi:2021pdp}.Therefore, Hypothetical coverage with template bank requires $10^{40}$  templates\cite{gair2004event}.
% Furthermore, non-local parameter degeneracy in the intrinsic parameter space leads to a multi-modal structure in the posterior distribution\cite{chua2022nonlocal}.

The task of an end-to-end data analysis process can be divided into three distinct steps~\cite{chua2022one}. (i) Detection: establishing the statistically significant presence of a GW signal in noisy detector data. This step can, for example, be performed relatively quickly and with a low false alarm rate using machine learning methods~\cite{zhang2022detecting}. (ii) Identification: mapping the detected signal (sufficiently) accurately to the source parameters of a (sufficiently) representative forward model which can guide the next step of parameter inference. Phenomenological waveforms~\cite{Wang2012xh} and harmonic search methods~\cite{babak2009algorithm} can both identify the phase evolution of the signal, and provide the information of the source parameters. Other methods have been applied to these two processes, such as the semicoherent method~\cite{gair2004event}, $\mathcal{F}$–statistic algorithms~\cite{wang2015first}, and time-frequency algorithms~\cite{gair2005detecting,wen2005detecting,Gair:2008ec}. (iii) Inference: estimating the Bayesian posterior probability of the actual source parameters, using for example, a Metropolis-Hastings search~\cite{Gair2008a,chua2022nonlocal}, Gaussian processes~\cite{Chua:2016jnd,Chua:2019wgs}, or the one-step likelihood function method~\cite{chua2022one}. Due to the algorithm's limited sampling efficiency and the waveform's computational speed, all inference processes are performed within the range of $\sim 100$ standard deviation  ($\sigma$) around the true value. Therefore, the signal identification process must already provide information about the parameters to control the inference process.

An EMRI signal is composed of many harmonics that are functions of the three fundamental orbital frequencies related to the radial $r$-motion, the azimuthal $\phi$-motion, and the polar $\theta$-motion, which are all evolving in time~\cite{AK_2004_PRD,chua2021rapid}. Non-local parameter degeneracy in the physical intrinsic space, defined by parameters such as the MBH mass $M$, MBH spin $a$, SOBH mass $\mu$, semi-major axis $p$, orbital eccentricity $e$, and orbital inclination $\iota$, typically leads to similar phase evolutions.
This means that even though the physical parameters may differ significantly, their phenomenological evolution can be very similar~\cite{babak2009algorithm,chua2021non}.  The one-step likelihood function method can break this multipeak structure but it only flattens these local peaks and thus does not effectively guide us into the signal neighborhood (the range of $\sim 100\,\sigma$)~\cite{chua2022one}.

In this work, we implemented an identification of an EMRI system, using simulated TianQin data as an example. 
In this work, the data processing is divided into three stages.
In the first stage, the relativistic waveform is used to search the whole parameter space, leading to a rough estimate for the parameters for the phenomenological waveform, which serves as a reference for the follow-up physical parameter search. In the second stage, a semicoherent phenomenological waveform search in the phenomenological parameter space is performed, which can efficiently identify the signal without the challenge of multipeak features in the physical parameter space. 
In the third stage, the intrinsic parameters of the EMRI are further refined within the parameter space bounded by the search results from the earlier stages.

% In this work, we have tried to inversion parameter information by using harmonic and phenomenological waveform search to identify 
% the dominant harmonic phase evolution.In order to efficiently find neighborhood signals in the full parameter space, the whole data processing is divided into three stages:In the first stage,harmonic mode search in the full parameter space and output the range of phenomenological waveform parameters and the physical parameters for subsequent searches.In the second stage,phenomenological waveform search in phenomenological parameters space,which greatly improved search efficiency.In the third stage, the intrinsic parameters of EMRIs are retrieved based on previous search results.

This paper is organized as follows. In Sec.~\ref{sec:background}, we describe the generation of the simulated data used. The approach to search for the signal's neighborhood is introduced in Sec.~\ref{sec:3stagesearch}. Sec.~\ref{sec:harmonicsearch} presents the method and results for the search with the physical harmonic waveform. The method and results for the phenomenological waveform detection are presented in Sec.~\ref{sec:analysisphenom}. In Sec.~\ref{sec:inversion}, we show the results of parameter inversion. Conclusions and discussions are provided in Sec.~\ref{sec:conclusion}. Throughout the paper, we use geometrical units where $G=c=1$.

\section{Data simulation}
\label{sec:background}
% key: introduce \ac{EMRI} formation and EMRIs waveform
%%%%这里需要一段话来把这里的几个小的章节串联起来；
%模拟数据的原则是尽可能地接近真实数据，
%模拟数据的构成
This section primarily presents the methods used to generate the simulated data, which is intended to replicate as closely as possible the characteristics of real data observed in future detections~\cite{MLDC_1_Arnaud_2007,Babak:2008aa,MockLISADataChallengeTaskForce:2009wir}. In order to verify the reliability of the principles underlying the entire signal identification process, we assume the simplest scenario. The simulated data $d = s + n$ is the corresponding Michelson streams, which are a combination of the GW signal $s$ and the detector noise $n$.

\subsection{EMRI waveform}

%__introduce modying waveform diffcult
Building EMRI waveforms is a challenging task for two main reasons. (i) The phase error of the waveform template needs to be as small as $\Delta \Phi \lesssim 1 / \rho$, where $\rho$ is the signal-to-noise ratio (SNR) of the source~\cite{amaro2011gravitational,katz2021fast}. To reach such an accuracy, it is required to consider the effects of the gravitational self-force caused by the small celestial body's own gravity~\cite{PT_Barack,self-force_Van}. (ii) The calculation time to generate waveform should be less than a second, because the typical stochastic sampling methods require  a large number of templates for likelihood evaluation. 
%For example, a semi-coherent search requires $\sim 10^{15}$ templates. 
Therefore, early data processing of EMRI signals used semirelativistic kludge waveforms such as AK~\cite{AK_2004_PRD} or AAK~\cite{AAK_Chua_2015_PRD}. Kludge waveforms trade accuracy for efficiency by means of a modular build and various computational approximations. Since these kludge waveforms do not consider the effects of the gravitational self-force, there will be a difference of tens to hundreds of radians relative to the real waveform on the radiation response timescale~\cite{osburn2016highly}.

Therefore, we use in this study the fully relativistic waveforms FEW to accurately identify the EMRI signal and track EMRI phase evolution. In the Schwarzschild eccentric condition, FEW can generate fast, accurate, and fully relativistic waveforms~\cite{chua2021rapid,katz2021fast}. This condition allows us to calculate the accurate phase evolution by considering the adiabatic-order gravitational self-force~\cite{van2018gravitational,hughes2021adiabatic}. Moreover, FEW uses order-reduction and deep-learning techniques to derive a global fit for harmonic modes of EMRIs so that the waveform can be generated in under $1\,{\rm s}$~\cite{chua2019reduced}. Using the Schwarzschild condition allows us to exclude various parameters. We can set the spin parameter of the MBH to be zero, $a=0$, so that the spacetime becomes spherically symmetric, which further allows us to remove the inclination parameter ($x_I=1$) as we can consider any orbit to be in the equatorial plane. Moreover, we can remove the polar phase $\Phi_\theta$ in this configuration.

EMRI waveforms can be represented by the complex time domain dimensionless strain $h(t)=h_{+}-i h_{\times}$, where $h_{+}$ and $h_{\times}$ are the normal transverse-traceless gravitational wave polarizations. The waveform $h(t)$ can then be written as~\cite{katz2021fast}
\begin{equation}
    h=\frac{\mu}{d_{L}} \sum_{l m n} A_{l m n}(t)_{-2} Y_{l m}(\theta, \phi) e^{-i \Phi_{m n}(t)},
\label{eq:EMRI_waveform}
\end{equation}
where $_{-2}Y_{l m}(\theta, \phi)$ are the %regular 
$-2$ spin-weighted spherical harmonics functions, $\theta$ is the source-frame polar viewing angle, $\phi$ is the source-frame azimuthal viewing angle, $\mu$ is the mass of the stellar-mass compact object, $d_L$ is the luminosity distance of  the source in the observer frame,  $l,m,n$ are the indices of the orbital angular momentum, the azimuth modes, and the radial modes, respectively.
The phase $\Phi_{mn} = m \Phi_{\varphi} + n \Phi_{r}$ represents the summation of decomposed phases for each given mode. It is noteworthy that the index for the polar phase $\Phi_\theta$ is k = 0.
In the search for GW signals, methods like the matched filtering are much more sensitive to the phase evolution than to the amplitude. Since the index $l$ only appears in the amplitude, while we are more interested in the phase, it would be convenient to absorb the $l-$dependency by summing over $l$.

% In matched filtering, the harmonic amplitude is estimated through a normalized template related to the harmonic phase, refer to Sec.\ref{sec:harmonic_principle} for more details.Therefore, it is difficult to independently detect the index $l$. We can sum over the index $l$ to obtain a simplified expression of the waveform}
%
\begin{equation}
    h=\frac{\mu}{d_{L}} \sum_{m n} A_{m n}(t)(\theta, \phi) e^{-i \Phi_{m n}(t)} .
\label{eq:EMRI_waveform_reduced}
\end{equation}

The adopted Schwarzschild eccentric waveform provides an accurate and fast calculation of EMRI waveforms required to verify the feasibility of the algorithm presented in this paper. In particular, when focusing on the sensitivity of detection algorithms to the orbital phase evolution. Therefore, we use these waveforms for the generation of the signal and for parameter extraction.
%%%--------------------------------------------------------------------------------%%%
%%% Revised on September 18 (end)

%%% Revised on September 19 (start)
%%%--------------------------------------------------------------------------------%%%
\subsection{Detection with TianQin}
% key: TianQin GW detector response

% ___introduce TianQin detector___
TianQin is a proposed spaceborne laser interferometer detector consisting of three identical drag-free satellites orbiting Earth in a nearly equilateral triangle~\cite{luo2016tianqin,mei2020tianqin,fan2020science}. The distance from the center of the Earth to each satellite is about $10^{5}\,\mathrm{km}$, which results in an armlength of about $1.73 \times 10^5\,\mathrm{km}$. Each satellite has a Kepler orbital period of about 3.64 days due to Earth's gravity and the detector's plane is oriented so that its normal vector points toward the reference source RX J0806+1527. The three satellites of the detectors are connected to each other by laser links to form Michelson interferometers.

% ___introduce TianQin response fuction___
The detected GW signal strain $h(t)$ can be described as a linear combination of the two GW polarizations ${h_{+,\times}}(t)$ modulated by the detector’s response
\begin{equation}
    h(t) = {F_ + }(\theta_{S} ,\phi_{S} ,\psi ){h_ + }(t) + {F_ \times }(\theta_{S} ,\phi_{S} ,\psi ){h_ \times }(t).
\end{equation}
Here $\theta_S$ and $\phi_S$ are the sky location of the source and $F_{+,\times}$ are the so-called antenna pattern functions
\begin{equation}
\begin{aligned}
    & F_{+}(t)=D_{+}(t, f) \cos (2 \psi)-D_{\times}(t, f) \sin (2 \psi), \\
    & F_{\times}(t)=D_{+}(t, f) \sin (2 \psi)+D_{\times}(t, f) \cos (2 \psi), 
\end{aligned}
\end{equation}
where the polarization angle $\psi$ is defined as
\begin{equation}
    \psi=\frac{\cos \theta_S \sin \theta_K \cos \left(\phi_S-\phi_K\right)-\sin \theta_S \cos \theta_K}{\sin \theta_K \sin \left(\phi_S-\phi_K\right)}.
\end{equation}
Here, $\theta_K$ and $\phi_K$ are the azimuthal and polar angles of the orbital angular momentum vector, respectively.

In the low-frequency regime ($f<f_{\ast}\approx 0.28\,{\rm Hz}$) $F_{+,\times}$ becomes independent of the GW's frequency $f$, and $D_{+,\times}$ can be analytically approximated as~\cite{Hu2018}
\begin{widetext}
\begin{subequations}
\begin{eqnarray}
  {D_+}(t;\theta_S,\phi_S) =&& \frac{{\sqrt 3 }}{{32}}\bigg(4\cos (2{\kappa _1})\Big(\big(3 - \cos(2\theta_S)\big)\cos{\bar \theta}\sin(2\phi_S - 2{\bar \phi})
    + 2\sin(\phi_S - {\bar \phi})\sin(2\theta_S)\sin({\bar \theta}) \Big)  \nonumber\\
  && -\sin(2{\kappa _1})\Big(3 + \cos(2\phi_S - 2{\bar \phi})\big(9 - \cos(2\theta_S)(3 + \cos(2{\bar \theta}))\big)
  - 6\cos(2{\bar \theta})\sin^2(\phi_S - {\bar \phi})  \nonumber\\
  &&+ 6\cos(2\theta_S)\sin^2({\bar \theta})
   + 4\cos(\phi_S - {\bar \phi})\sin(2\theta_S)\sin(2{\bar \theta})\Big)\bigg),  \\
  {D_\times}(t;\theta_S,\phi_S) =&& \frac{{\sqrt 3 }}{8}\bigg(-4\cos(2{\kappa _1})\Big(\cos(2\phi_S - 2{\bar \phi})\sin(\theta_S)\cos({\bar \theta})
   + \cos(\phi_S - {\bar \phi})\cos(\theta_S)\sin({\bar \theta})\Big) \nonumber\\
   &&+ \sin(2{\kappa _1}) \Big(-\sin(\theta_S)\big(3 + \cos(2{\bar \theta})\big)\sin(2\phi_S - 2{\bar \phi})
   - 2\sin(\phi_S - {\bar \phi})\cos(\theta_S)\sin(2{\bar \theta})\Big)\bigg).
\end{eqnarray}
\end{subequations}
\end{widetext}
Here $\kappa_{1} = 2\pi f_{\rm{sc}}t + \kappa _0$, $f_{\rm{sc}}=1/(3.64$ day), and $\kappa _0$ is the constant phase determined by the setup of the satellites' coordinates (see \cite{Hu2018} for details). Moreover, $(\bar{\theta} = 1.65, \bar{\phi} = 2.10)$ are the colatitude and
longitude of the reference source RX J0806+1527~\cite{Luo2015}.

A correction term is further added to account for the Doppler effect induced by the orbital motion of the TianQin satellites~\cite{liu2020science,fan2020science}
\begin{equation}
    \Phi^D(t) = 2\pi\nu(t)R\sin(\theta_S)\cos(\phi(t)-\phi_S),    
\end{equation}
where $2\nu(t)$ is the frequency of the GW signal, $R=1\,\mathrm{AU}$, $\phi(t)=\phi_0+2 \pi t / T$ with $T=1\,\mathrm{yr}$ the Earth's orbital period around the Sun, and $\bar{\phi}_0$ is the initial location of TianQin at the time $t=0$.

We simulate TianQin's noise assuming it is Gaussian and stationary. It is then encoded in the following sensitivity curve
\begin{equation}
\begin{aligned}
    S_n(f)=\frac{1}{L^2}\Bigg[\frac{4 S_a}{(2 \pi f)^4}\Big( & 1+\frac{10^{-4} \mathrm{H} z}{f}\Big)+S_x\Bigg] \\
    \times & {\left[1+0.6\left(\frac{f}{f_*}\right)^2\right], }
\end{aligned}
\label{eq:noise_curve}
\end{equation}
where $S_a^{1/2}=1\times10^{-15}$m s$^{-2}/$Hz$^{1/2}$ characterizes the residual acceleration on a test mass playing the role of an inertial reference, $S_x^{1/2}=1\times10^{-12}$m/Hz$^{1/2}$ characterizes the one-way noise of the displacement measurement with inter-satellite laser interferometry, and $f_*=c/(2\pi L)$ is the transfer frequency~\cite{Luo2015}, with $L$ being the armlength.

Using the one-sided PSD given in Eq.~(\ref{eq:noise_curve}), the SNR can be defined as
\begin{equation}
    \rho=(d \mid h)^{1 / 2}=2\left[\int_0^{\infty} \frac{\tilde{d}(f) \tilde{h}^*(f)}{S_n(f)} {\rm d} f\right]^{1 / 2},
\end{equation}
where $(\cdot|\cdot)$ denotes the noise-weighted inner product defined on the right side of the equation, and $\tilde{d}(f)$ and $\tilde{h}(f)$ are the data and the template in the frequency domain, respectively.

\subsection{Data}

We consider a GW signal originating from an EMRI formed by a SOBH orbiting a Schwarzschild MBH. We assume for the observation time of the data $0.5\,{\rm yr}$ and set the EMRI parameters as follows: the mass of the MBH is $M=10^6\,{\rm M_\odot}$, the mass of the SOBH is  $\mu =10\,{\rm M_\odot}$, the initial orbital eccentricity is $e_0=0.2$, and the initial semilatus rectum is $p_0=8.5\,M$. We further set the sky position of the source to $(\theta_S,\phi_S)=(\pi/4,\pi/4)$, the polarization angle to $\psi= 1.125$, and the initial phases to $(\Phi_{\phi,0},\Phi_{r,0}) = (1.0,3.0)$ where they can be random in the interval $[0,\pi]$. The plunge of the SOBH in the MBH occurs around $0.44\,{\rm yr}$ after the detection begins. The simulated data is then obtained by applying TianQin's Michelson response to the EMRI waveform.

% The plunge time $t_{p}$ of the signal is 0.394 year,which is Random in [0.3,0.5] year.
% We set the parameters of the EMRI by the following:
% the mass of the MBH $M=10^6M_\odot$, the mass of the compact  object
% (stellar mass BH)  $m=10M_\odot$, the initial orbital eccentricity
% $e=0.2$,the semi-latus rectum $p=8.4364 M$ is calculated by $M,m,e,t_{p}$,the sky position of the
% source $(\theta^S,\phi^S)=(\pi/4,\pi/4)$,
% the polarization angle $\psi= 1.125$.
% ,the initial phase $(\Phi_{\phi0},\Phi_{r0}) = (4.0977,2.7959)$,which is Random in $[0,\pi]$, In the Schwarzschild eccentric condition,the MBH's spin $a$ and the initial phase $\Phi_{\theta}$ are ignored,and the inclination angle
% $\iota=0$.

\begin{figure}[htbp]
\centering
\includegraphics[width=0.5\textwidth]{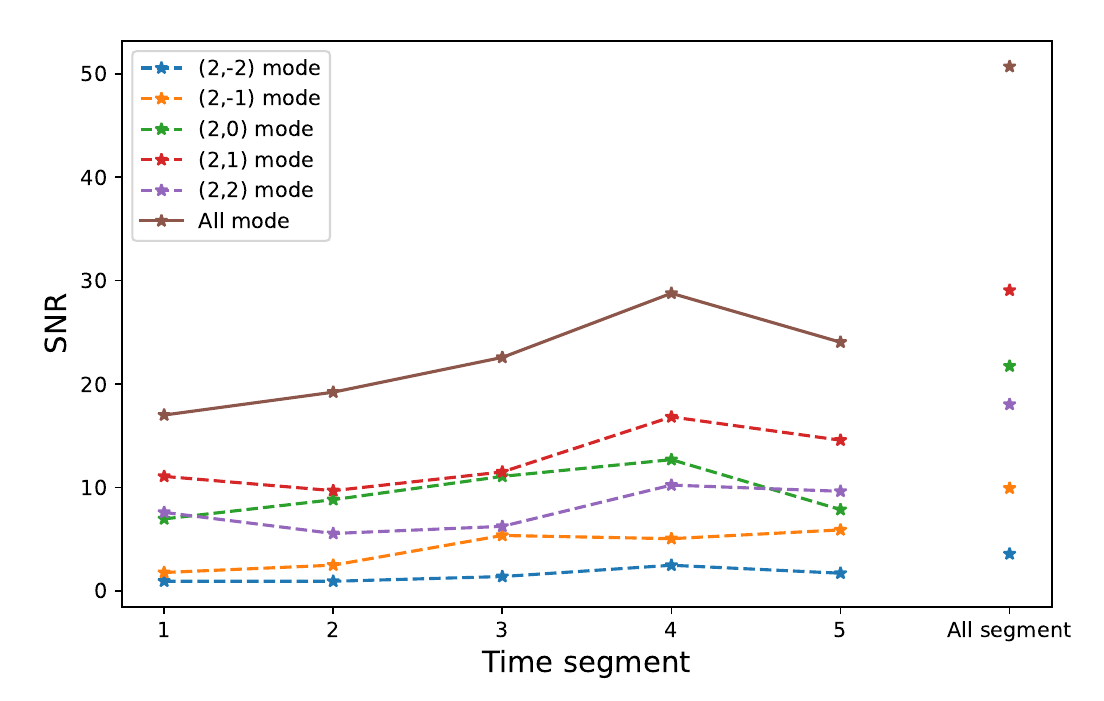}
    \caption{The SNR of the five dominant harmonic modes for the entire observation time and the time segments considered.}
\label{fig:signal_SNRI}
\end{figure}

The total SNR of the source is 50 and the SNR of the dominant harmonic modes ($m=2$, $n\in[-2,2]$~\cite{chua2021non}) is shown in Fig.~\ref{fig:signal_SNRI}. We note that the search algorithm crucially depends on the SNR of the source and in particular on the SNR of the dominant modes. Although the SNR of the dominant modes in the separate time segments is relatively low, precise measurements can still be achieved through the search with phenomenological waveforms. %The data described above is used for all searches. We only consider a single EMRI signal excluding several parameters but we believe that this is sufficient to verify the principles of the algorithm. See Section~\ref{sec:conclusion} for a detailed discussion.

% To validate the validity of the entire detection algorithm, 
% we will use the simulated data set with 6 months of observation time.
% The data $d = s + n$ is the corresponding Michelson streams, which is combined a GW signal $s$ and the detector noise $n$.

% \begin{table}[htp!]
% \centering
% \begin{tabular}{|c|c|c|c|c|c|c|}
% \hline  % 在表格最上方绘制横线
% (m,n) &  (2,-2) & (2,-1) &  (2,0) &  (2,1) &  (2,2)&	 \thead{Sum of \\ all modes} \\
% \hline  %在第一行和第二行之间绘制横线				
% $[0.0,0.5]\,{\rm yr}$&3.59&9.94&21.74&29.07&18.04& 50.7 \\   
% \hline 
% $[0.0,0.1]\,{\rm yr}$ &0.95&1.79&6.97&11.08&7.60& 17.01 \\   
% \hline 
% $[0.1,0.2]\,{\rm yr}$ &0.94&2.50&8.83&9.70&5.56& 19.22 \\   
% \hline
% $[0.2,0.3]\,{\rm yr}$ &1.40&5.37&11.08&11.51&6.25& 22.58 \\  
% \hline
% $[0.3,0.4]\,{\rm yr}$ &2.49&5.06&12.70&16.83&10.24& 28.78 \\ 
% \hline
% $[0.4,0.5]\,{\rm yr}$ &1.72&5.91&7.87&14.58&9.64& 24.05 \\   
% \hline  
% \end{tabular}
% \caption{The SNR of the five dominant harmonic modes for the entire observation time and the time segments considered.}
% \label{table:signal_SNRI}
% \end{table}

%%%--------------------------------------------------------------------------------%%%
%%% Revised on September 19 (end)

% \begin{table*}[htp!]
% \centering
% \caption{The SNR of five harmonic modes with not noise}
% \begin{tabular}{|c|c|c|c|c|c|c|}
% \hline  % 在表格最上方绘制横线
% (l,m,n) &  (2, 2, -2) & (2, 2, -1) &  (2, 2, 0) &  (2, 2, 1) &  (2, 2, 2)&	 \thead{Sum all \\ modes} \\
% \hline  %在第一行和第二行之间绘制横线				
% $[0.0,0.5]$yr&	2.03&9.89&21.99&29.14&17.75	&50.7\\
% \hline  
% $[0.0,0.1]$yr&	0.14&	2.29&	7.65&	9.57&	6.16&17.01\\
% \hline  
% $[0.1,0.2]$yr&	0.19&	2.85&	8.68&	11.01&	6.66&19.22\\
% \hline  
% $[0.2,0.3]$yr&	0.31&	3.61&	10.59&	12.77&	7.61&22.58\\
% \hline  
% $[0.3,0.4]$yr&	0.72&	5.66&	13.07&	16.50&	9.53&28.78\\
% \hline  
% $[0.4,0.5]$yr&	1.85&	6.28&	8.18&	14.17&	9.15&24.05\\
% \hline  
% \end{tabular}
% \label{table:signal_SNRI}
% \end{table*}

%%% Revised on September 20 (start)
%%%--------------------------------------------------------------------------------%%%
\section{Three stage signal search} % Feel free to change the title if you have a better one
\label{sec:3stagesearch}

As described in the introduction, we present in this work an approach to search for the neighborhood of the signal in three stages. The first step, where a search using physical harmonic waveforms in the whole parameter space is implemented to find a rough range of the phenomenological waveform parameters and the physical parameters that are used in the subsequent search is introduced in Sec.~\ref{sec:harmonicsearch}. In Sec.~\ref{sec:analysisphenom}, we present the semicoherent phenomenological waveform search, which is used to improve the search efficiency. The third stage, where the intrinsic parameters of the EMRI are retrieved based on the previous search results is presented in Sec.~\ref{sec:inversion}.

\subsection{Harmonic mode search}
\label{sec:harmonicsearch}

In this section, we begin the initial search for signals using physical harmonic waveforms. Despite the relatively time-consuming process of generating physical waveforms compared to phenomenological waveforms, we still choose to employ physical waveforms as the first stage in our search for the following two reasons: First, physical waveforms are directly correlated with the parameters of the EMRI system, allowing us to approximate the range of signal parameters as long as the templates match the signal. Second, the harmonic information within the physical waveforms can directly guide the search for phenomenological waveforms.%, thereby reducing the parameter space.

% The harmonic mode search method here is similar to \cite{babak2009}.
% In the first stage,harmonic mode search in the full parameter space and output the range of phenomenological waveform parameters and the physical parameters for subsequent searches. \ato{I find to short to understand what is being done.}

% In the first stage,we aim to provide the priori information of the phenomenological waveform parameters in Sec.\ref{sec:analysisphenom} and the physical parameters
% in Sec.\ref{sec:inversion}.

\subsubsection{Detection principle}
\label{sec:harmonic_principle}

The advantage of the harmonic mode search is that it allows for data matching by the different harmonic modes of the EMRI signal. It is difficult to find a template where all the harmonics exactly match the corresponding harmonics of the signal. Therefore, we search for templates where the harmonics match the harmonics of the signal independent from the indices of the harmonics, {\it i.e.}, we also allow for matches where the indices of the harmonics of the template do not correspond to the indices of the harmonics of the signal. The effectiveness of this method depends on the harmonics SNR, with higher SNRs yielding better results. During the search, we generate many templates and give the possible parameter range of the signal according to the distribution of the SNR of the corresponding templates.

%Similar to Ref.~\cite{babak2009}, 
The initial search involves the maximization of the plunge time and the initial phase $\Phi$. The maximization of the plunge time is similar to the one used in Ref.~\cite{babak2008building}, where the correlation of the template with the data is computed (instead of using the inner product). The correlation $C_{h, s}(\tau)$ is defined as~\cite{babak2008building}
\begin{equation}
    C_{h, s}(\tau)=2 \int \frac{\tilde{s}(f) \tilde{h}^*(f)}{S_n(f)} e^{i 2 \pi f \tau} {\rm d} f
\end{equation}
where $\tau$ is the time lag between the signal and the template.

We introduce here briefly the maximization of the initial phase, and interested readers are suggested to check Ref.~\cite{babak2009} for more details. Each harmonic of the waveform can be approximated as
\begin{equation}
    h = A_{\rm cons} A(t) \cos(\Phi_0+ \tilde{\Phi}(t))
\end{equation}
where $A_{\rm cons}$ and $A(t)$ are the amplitudes of the constant parts and the time-related parts, respectively. Therefore, $h$ can be decomposed as
\begin{equation}
    h = \cos{\Phi_0} h(0) - \sin{\Phi_0} h(\pi/2),
\end{equation}
where $h(0)$ is the value of the harmonic $h$ taken at zero initial phase. Omitting all cross harmonic terms, $a_0 = A_{\rm cons}\cos{\Phi_0}$ and $a_1 = A_{\rm cons}\sin{\Phi_0}$ can be obtained from the following ratios of inner products (similar to $\mathcal{F}$-statistics)
\begin{equation}
    a_{0} = \frac{(d|h(0))}{({h}|{h})}, \quad
    a_{1} = \frac{(d|h(\pi/2))}{({h}|{h})}.
\label{acoef}
\end{equation}
Using $a_0$ and $a_1$, we then get the following maximum likelihood estimators for the
amplitude and phase of the harmonics
\begin{equation}
    \Phi_0 = \arctan{\left(\frac{a_{1}}{a_{0}}\right)}, \quad
    A_{\rm cons} = \sqrt{a_{0}^2 + a_{1}^2}.
\label{physMax}
\end{equation}
Notice that for each harmonic search, we are more interested in $A_{\rm cons}$ since it is equivalent to the harmonic's SNR. 

We start the search using a random template bank with parameter distributions as listed in Table~\ref{tab:AKW-parameters}. To optimize over the plunge time, each set of parameters in the template bank is used to generate the waveform for the last half year of the inspiral before plunge $t_{p} = 0.5\,{\rm yr}$. After obtaining the best fit of the plunge time, we use the harmonic waveform $m=2$, $n\in[-2,2]$ to match the signal respectively. We calculate  
the harmonic’s SNR for each harmonic by optimizing over the initial phase using Eq.~(\ref{physMax}). 
From this result, we obtain the distribution of the harmonic's SNR after completing the search for a large enough number of sets of parameters. The mean $\mu_{\theta}$ and the standard deviation $\sigma_{\theta}$ of the parameter distribution is then obtained by setting an appropriate threshold for the SNR to set the parameter range $[\mu_{\theta} - \sigma_{\theta},\mu_{\theta} + \sigma_{\theta}]$.

\begin{table*}[htbp]
\centering
\begin{ruledtabular}
\begin{tabular}%{|c|c|l|}
{p{0.06\textwidth}<{\centering}%
p{0.46\textwidth}<{\centering}%
p{0.42\textwidth}<{\centering}}
    \textbf{Symbol} & \textbf{Physical Meaning} & \textbf{ distribution}  \\
    \hline  
    $M$ &  MBH mass & uniform in $\ln{M}$ over $[10^4,10^7]\,{\rm M_{\odot}}$\\ 
    $\mu$   & SOBH mass & uniform in $[5-15]\,{\rm M_{\odot}}$ \\ 
    $e_{0}$  & initial orbital eccentricity at $t_{0}$ & uniform in $[0,0.7]$ \\
    $p_{0}$  & semilatus rectum at $t_{0}$ & depends on $[t_p,M,\mu,e_0]$ \\
    $\theta_S$ & polar angle to the source & uniform in $[0,2\pi]$\\ 
    $\phi_S$  & azimuthal angle to source & uniform in $[0,\pi]$\\
    $\psi$ & polarization angle & uniform in $[0,2\pi]$ \\
    $\Phi_{\varphi}$ & azimuth phase & uniform in $[0,2\pi]$ \\
    $\Phi_{r}$ & radial phase & uniform in $[0,2\pi]$ \\
    $\theta_K$ & polar angle of the source's angular momentum  & uniform in $[0,2\pi]$ \\ 
    $\phi_K$ & azimuthal angle of the source's angular momentum & cos($\phi_K$) is uniform in $[-1,1]$ \\ 
    $D_L$  & luminosity distance of the source & 1 Gpc \\
\end{tabular}
\end{ruledtabular}
\caption{Meaning of the physical parameters in FEW and their distribution considered. In the Schwarzschild eccentric condition, the MBH's spin vanishes, $a=0$, resulting in a fixed inclination angle $\iota=0$ and the initial phase $\Phi_{\theta}$ can be ignored.}
\label{tab:AKW-parameters} 
\end{table*}

\subsubsection{Harmonic detection results}

\begin{figure*}[htbp]
\centering
\includegraphics[width=1\textwidth]{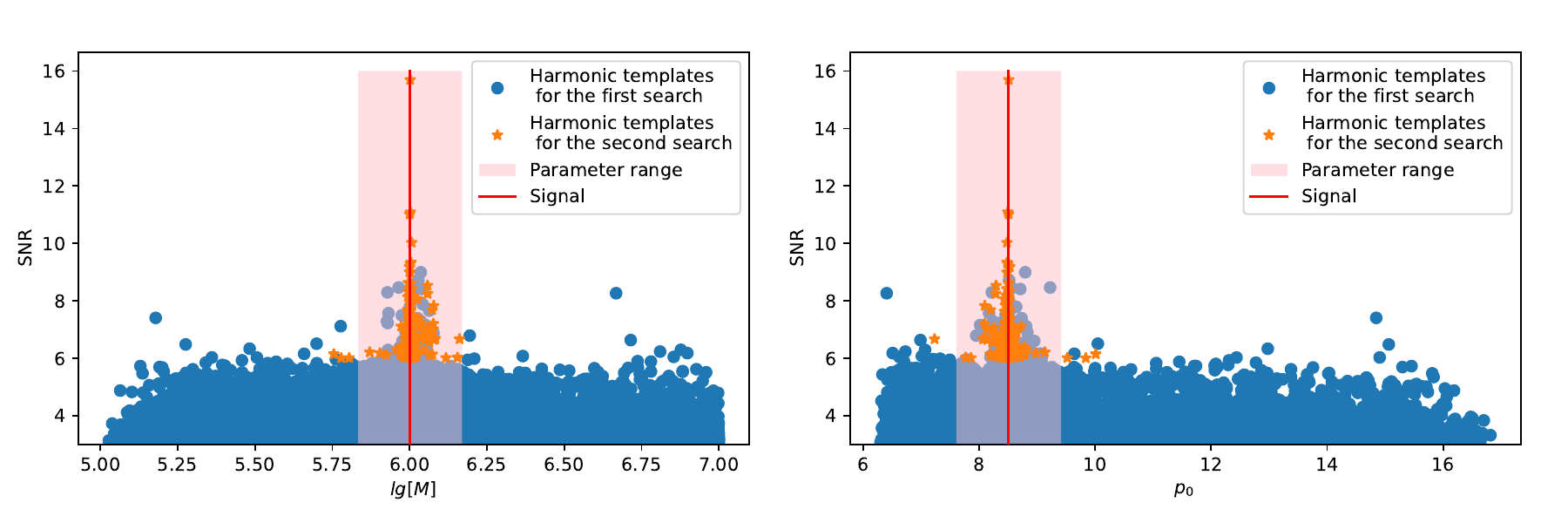}
    \caption{The distribution of the harmonic's SNR for the parameters $M$ and $p_0$. We select only the strongest harmonic mode for each template. Templates with a high harmonic SNR cluster near the signal.}
    \label{fig:M_p_distribution}
\end{figure*}

In this stage, we use as many templates as possible keeping the calculation time acceptable. As we show later, a total of 600,000 templates were utilized for the search, implemented on a cluster with 100 cores. FEW generates an EMRI waveform in $\sim 5\,{\rm s}$, thus the calculation time for the entire search is of $1-2$ days. A good search result is given when five ($n\in[-2,2]$) harmonics of a template match with five harmonics of the signal. However, finding a template with multiple harmonic modes matching the signal is challenging. Therefore, we select the harmonic mode that best matches the signal to present our search results. The distribution of the harmonic's SNR for the parameters $\log(M)$ and $p_0$ of these templates are shown in Fig.~\ref{fig:M_p_distribution}. As can be seen, a large number of templates with high harmonic SNR accumulate near the signal.

Next, we set the threshold value and boundaries for the search results to obtain the range of physical parameters. We select the top $\sim 0.01\,\%$ templates with the highest SNR corresponding to a SNR threshold of around 6. For the first search, we use approximately 500,000 templates. Among them are around 50,000 templates with ${\rm SNR}>3$ and only 80 templates with ${\rm SNR}>6$. We see in Fig.~\ref{fig:M_p_distribution} that there are a few templates with ${\rm SNR}>6$ further away from the signal but most of the high SNR templates cluster around the signal. This phenomenon is particularly evident for the parameters $M$ and $p_{0}$. To obtain the boundaries for the physical parameters, we calculate the mean and standard deviation for the distribution of the templates. The mean value and the standard deviation for $\log_{10}(M/{\rm M_\odot})$ are $6.02$ and $0.29$, respectively, while the mean value and the standard deviation for $p_{0}$ are $8.75\,M$ and $1.52\,M$, respectively. We focus on the points locate in the boundaries $[\mu_\theta-\sigma_{\theta},\mu_\theta+\sigma_{\theta}]$, and update the parameters to  $[5.73,6.31]$ and $[7.23,10.27]\,M$ for the ranges of $\log_{10}(M/{\rm M_\odot})$ and $p_{0}$, respectively.

The initial search provides a rough indication of the source parameter.
We refine the search by performing a second round of search in the zoomed-in range. 
%If a template exceeds the threshold, it indicates that the harmonic phase matches the signal's harmonic phase well. To capture more templates matching the signal's harmonics,
In the second round, we expand the boundaries obtained from the initial search by 10\,\%. %and conduct a second search. 
For this search, we use approximately 100,000 templates, where we get 187 templates with a SNR greater than 6. We combine the results from this search and the previous search, and obtained a total of 267 templates with ${\rm SNR}>6$. The mean value and the standard deviation we get from the combined distribution for $\log_{10}(M/{\rm M_\odot})$ are $6.01$ and $0.16$, respectively, while the mean value and the standard deviation for $p_{0}$ are $8.58\,M$ and $0.90\,M$, respectively. Therefore, we obtain the final parameter range $[5.85, 6.17]$ and $[7.68, 9.48]\,M$ for $\log_{10}(M/{\rm M_\odot})$ and $p_{0}$, respectively.

\begin{figure*}[htbp]
\centering
\includegraphics[width=1\textwidth]{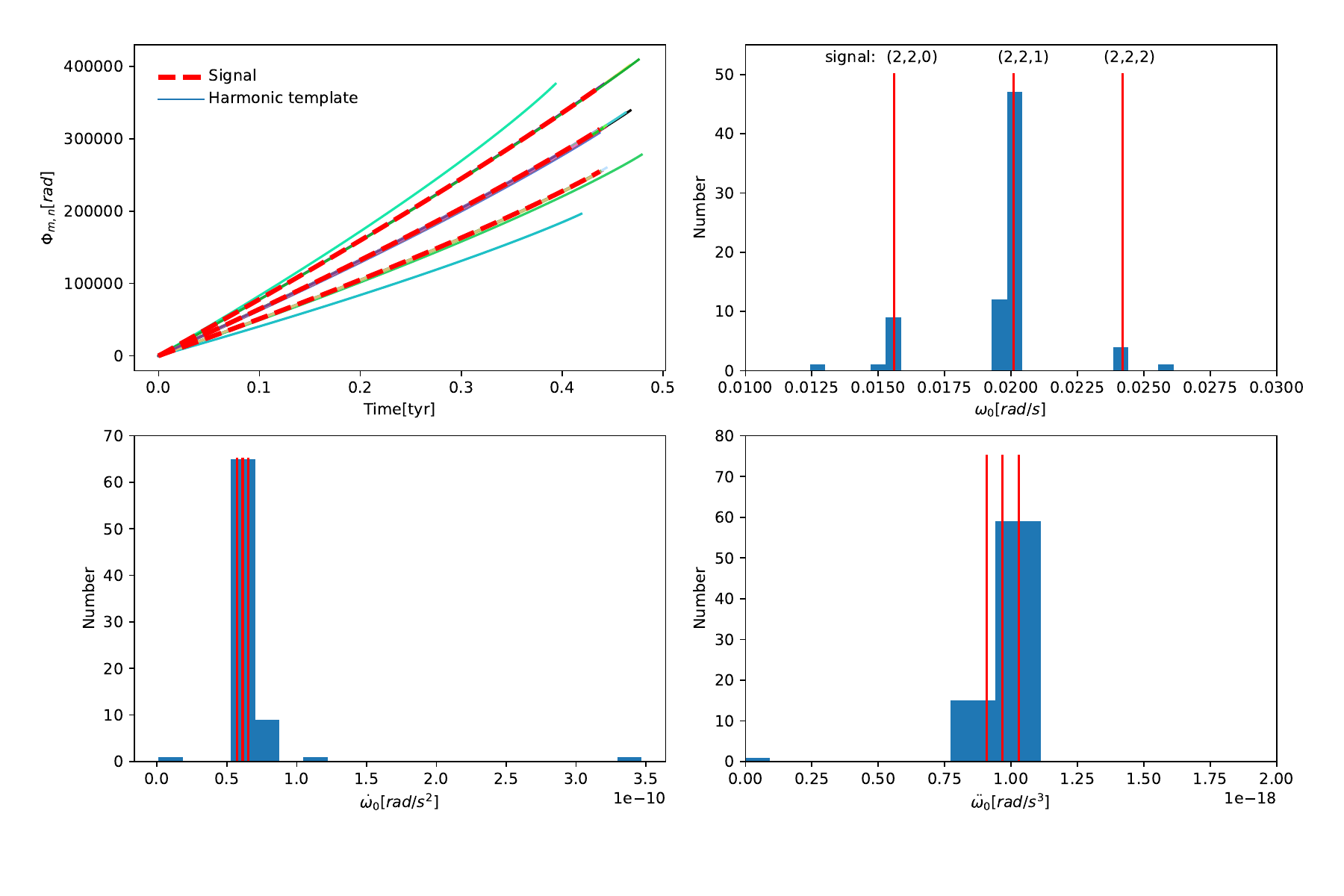}
    \caption{The phase evolution distribution of high SNR physical harmonic templates. The top-left panel illustrates the phase evolution for all high SNR templates.%, with each template selectively featuring the strongest harmonic mode. 
    The remaining panels show the phenomenological parameter distributions of $\omega_{0}$, $\dot{\omega_{0}}$, and $\ddot{\omega_{0}}$ (c.f. Sec.~\ref{sec:phenomwf}) in the initial time segment of $0.1\,{\rm yr}$, respectively.}
\label{fig:phidistribution}
\end{figure*}

In Fig.~\ref{fig:M_p_distribution}, results from the first coarse search are labeled in blue points and the pink-shaded region.
The orange stars represent the refined search results, and the injected value is indicated by the red vertical line.
After the refined search, the denser templates allow us to increase the SNR threshold, which we used value of 7, to concentrate on better-fit templates.
In the top left panel of Fig.~\ref{fig:phidistribution}, we present the phase evolution of the harmonics for all templates with ${\rm SNR}>7$, while the harmonics of the injected signal are represented with red dashed lines. We can clearly see that all templates with ${\rm SNR}>7$ have at least one harmonic whose phase evolution tracks the signal closely, and all phase evolutions display polynomial features.

\subsection{Semicoherent search with a phenomenological waveform}
\label{sec:analysisphenom}

Using the physical waveforms to search is very time-consuming, especially considering the multipeak nature. 
Hence we use a phenomenological waveform for the next search stage. The top left panel of Fig.~\ref{fig:phidistribution} indicates that polynomial functions might be sufficient to describe the phase evolution. %This stage improves the efficiency of the search and gives the posterior parameter distribution for the phenomenological parameters. 
Here, a semicoherent search is employed, where we divide the data into time segments of $0.1\,{\rm yr}$ and infer the phenomenological parameters for each time segment. 
Before discussing the details and results of the search in this stage, we introduce some basic concepts of the phenomenological waveforms we use in the following section.

%%%--------------------------------------------------------------------------------%%%
%%% Revised on September 20 (end)

% \sout{the information of physical parameters by parameter inversion in \ref{sec:inversion}.}

% In order to further improve the search efficiency and avoid the degeneracy between intrinsic parameters,we use a phenomenological waveform to do the next search.

%%% Revised on September 21 (start)
%%%--------------------------------------------------------------------------------%%%

\begin{figure}[!htbp]
\centering
\includegraphics[width=0.55\textwidth]{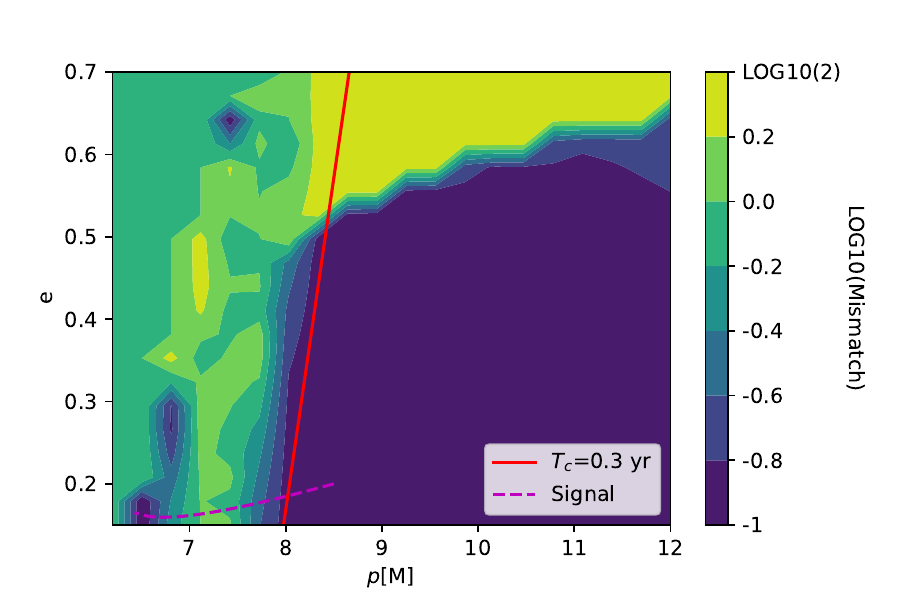}
\caption{The mismatch between the phenomenological and physical waveforms with different $p$ and $e$. The purple dashed line indicates the $p-e$ evolution of the signal from (8.5,0.2) to (6.43, 0.165). The red solid line indicates where the evolution time from $(p,e)$ to the plunge is $0.3\,{\rm yr}$, points on the right side being longer than $0.3\,{\rm yr}$.}
% \Yecq{The label of the colorbar in this picture is wrong and is being modified.}}
\label{fig:p_e_mismatch}
\end{figure}

\subsubsection{Phenomenological waveform}
\label{sec:phenomwf}

% __introduce why can ues phenomenological waveforms___
The key idea of searching for signals using a phenomenological waveform is that 
the evolution of the EMRI's orbital frequency is relatively smooth in the early stage of the inspiral. In a short period, the phase $\Phi_{m n}(t)$ can be Taylor expanded as~\cite{wang2012extreme}
\begin{equation}
\begin{aligned}
    \Phi_{m n}(t)  =& \Phi_{m n}\left(t_0\right)+\omega_{m, n}\left(t_0\right)\left(t-t_0\right) +\frac{1}{2} \dot{\omega}_{m, n}\left(t_0\right) \\
    &\times\left(t-t_0\right)^2+\frac{1}{3} \ddot{\omega}_{m, n}\left(t_0\right)\left(t-t_0\right)^3 + O(4),
\label{eq:phase_fitting}
\end{aligned}
\end{equation}
where we define the angular frequency $\omega_{m,n} =: m\omega_{\phi} + n\omega_{r}$, and $\dot{\omega}_{m, n}$ and $\ddot{\omega}_{m, n} $ are its first and second time-derivatives, respectively. The amplitudes in GWs $A_{mn}$ evolve even smoother than the phase over an extended period of time. Moreover, detection is more sensitive to a mismatch in the phase than in the amplitude. Therefore, for simplicity, we ignore the time evolution of the amplitudes and treat all of them as constants~\cite{wang2012extreme,hughes2021adiabatic}. 
The Doppler effect induced by the orbital motion of the detectors is also smoother than the phase over an extended time period and hence, we can also ignore it in the template.

%__introduce (p,e) mismatch ___这里没有详细介绍，每个点都是选取的0.1年的时长，进行计算的
To check  the similarity of the phenomenological waveform to the physical waveform, we calculate their mismatch, which is defined as
\begin{equation}
    {\mathrm Mismatch} =1 - \frac{(a \mid b)}{\sqrt{(a \mid a)(b \mid b)}},
\label{eq:overlap}
\end{equation}
where a and b are the phenomenological waveform and the physical waveform, respectively. 
 We demonstrate the efficacy of the phenomenological waveform by showing the mismatch between the physical waveform in Fig. ~\ref{fig:p_e_mismatch}. Each point in the figure represents a set of physical parameters, and all other parameters are fixed to the same values, such as $M =10^{6}\,\rm M_{\odot}$, $\mu = 10\,\rm M_{\odot}$. 
Then we let this parameter set evolve for the initial 0.1 year, obtaining the phase evolution and the physical waveform.
The mismatch of the best-fit phenomenological waveform and the physical waveform is calculated with Eq. \ref{eq:phase_fitting}. 
For better intuition, we indicate the $p-e$-evolution of a signal from (8.5,0.2) to (6.43, 0.165) with the dashed line.
The red solid line marks where the time from $(p,e)$ to plunge is $0.3\,{\rm yr}$, {\it i.e.}, a source with parameters on the right side from the red line needs longer than $0.3\,{\rm yr}$ to plunge. We can observe that regions to the right side of the red line show good consistency between the phenomenological waveform and the physical waveform. Considering shorter time segments for the semicoherent search can further increase the applicable range for phenomenological waveform~\cite{gair2004event}. 

\subsubsection{Template search}

% ___introduction the aim of Template _____

% In this stage, we use similar to semi-coherent searching, and divided the data into five segments, each with a duration of 0.1 years.

At this stage, a semicoherent approach is adopted to search with the phenomenological waveforms.
Semicoherent methods relax the stringent requirements on the phase accuracy of the models, combined with the usage of simpler waveforms like phenomenological waveforms for the search, the total computational time can be hugely compressed \cite{gair2004event}. In exchange, the semicoherent likelihood leads to wider posterior distributions, particularly on those parameters that strongly influence the phase of the GW signal. Semicoherent methods are widely used in searches for continuous waves in LIGO/Virgo data~\cite{riles2023searches,prix2012search}. Here, we divided the data into five segments, each with a duration of $0.1\,{\rm yr}$. Notice that the validity of the phenomenological waveform drops significantly when the system approaches merge, in practise only the results from the first three segments were used. 

In each segment, we search using three different harmonic modes of the phenomenological waveforms.
We sample the posterior distribution of the phenomenological parameters for each harmonic mode within each segment. The posterior probability distribution of the parameters $p(\theta|d)$ can be obtained using a Bayesian analysis
\begin{equation}
    p(\theta|d) = \frac{p(\theta)p(d|\theta)}{p(d)}
\end{equation}
where $p(\theta)$ is the prior probability distribution of the parameters, $p(d|\theta)$ is the likelihood, and $p(d)$ is the evidence. The evidence is a normalization factor that is independent of the parameters and hence we can ignore it here. The standard Bayesian (log-)likelihood ratio function of the source parameters $\theta$, is given by
\begin{equation}
\Lambda(\theta) := (d|h) - \frac{1}{2}(h|h).
\end{equation}

Next, we discuss how to select the prior for the parameters of a harmonic mode within each time segment. Based on the search results using physical waveforms in Sec.~\ref{sec:harmonicsearch}, we can obtain harmonic phases that match the signal. We divide these phases into segments of $0.1\,{\rm yr}$ each and fit them using Eq.~(\ref{eq:phase_fitting}). The upper right and the two lower subplots in Fig.~\ref{fig:phidistribution} show the distributions of the three phenomenological parameters $\omega_{0}$, $\dot{\omega}_{0}$, and $\ddot{\omega}_{0}$ for the harmonic phases in the first time segment, respectively. The distribution of $\omega_{0}$ shows three different harmonic modes. Despite the absence of evident harmonic structures in the parameter distributions of $\dot{\omega}_{0}$ and $\ddot{\omega}_{0}$, they can still provide a narrow parameter range. We summarize the prior parameter ranges of $\omega_{0}$, $\dot{\omega}_{0}$, and $\ddot{\omega}_{0}$ for the three main harmonic modes in the different time segments
in Table~\ref{table:fittingparameters}. Furthermore, we refer to Table~\ref{tab:AKW-parameters} for the selection of the priors for the extrinsic parameters.

\begin{table*}[htbp!]
\centering
\begin{tabular}{|c|c|c|c|c|c|c|c|c|}
\hline  % 在表格最上方绘制横线
\multirow{2}*{Time segment} & \multirow{2}*{modes} &  \multicolumn{2}{|c|}{$\omega_0$}&  \multicolumn{2}{|c|}{$\dot{\omega}_0$} &  \multicolumn{2}{|c|}{$\ddot{\omega}_0$}\\
\cline{3-8}
& & prior& signal&prior& signal&prior& signal\\
\hline
\multirow{3}*{1-segment} & (2,0) & [0.015,0.017] & 0.0158 & [0,$1 \times 10^{-10}$] & 5.71$\times 10^{-11}$ & [0,2$\times 10^{-18}$] & 1.03$\times 10^{-18}$ \\
\cline{2-8}
& (2,1) &[0.019,0.021]&0.0201 &[0,$1 \times 10^{-10}$] &6.13$\times 10^{-11}$ & [0,2$\times 10^{-18}$] & 9.68$\times 10^{-19}$ \\
\cline{2-8}
& (2,2) &[0.024,0.026]& 0.0244 &[0,$1 \times 10^{-10}$] &6.55$\times 10^{-11}$ & [0,2$\times 10^{-18}$]  &9.08$\times 10^{-19}$ \\
\hline
\multirow{3}*{2-segment} & (2,0) & [0.016,0.018]& 0.0166 &[0,$1 \times 10^{-10}$] &7.11$\times 10^{-11}$ & [0,3$\times 10^{-18}$] &  1.73$\times 10^{-18}$\\
\cline{2-8}
& (2,1) & [0.020,0.022]&0.021 &[0,$1 \times 10^{-10}$] & 7.46$\times 10^{-11}$ &[0,3$\times 10^{-18}$]& 1.56$\times 10^{-18}$ \\
\cline{2-8}
& (2,2) &[0.024,0.026]&0.0253 &[0,$1 \times 10^{-10}$] &7.80$\times 10^{-11}$ &[0,3$\times 10^{-18}$] &1.39$\times 10^{-18}$ \\
\hline
\multirow{3}*{3-segment}& (2,0) &[0.016,0.018]&0.0177 &[-$1 \times 10^{-10}$,$1 \times 10^{-10}$]  &9.32$\times 10^{-11}$ & [0,1$\times 10^{-16}$] & 3.57$\times 10^{-18}$\\
\cline{2-8}
& (2,1) & [0.019,0.024] & 0.022& [-$1 \times 10^{-10}$,$1 \times 10^{-10}$] &9.49$\times 10^{-11}$ & [0,1$\times 10^{-16}$]&  3.01$\times 10^{-18}$ \\
\cline{2-8}
& (2,2) & [0.023,0.027] & 0.026 & [-$1 \times 10^{-10}$,$1 \times 10^{-10}$] & 9.66$\times 10^{-11}$ & [0,1$\times 10^{-16}$]& 2.46$\times 10^{-18}$ \\
\hline
% 4-segment & (2,1)  & [0.018,0.029] & 0.023 &[-5e-9,5e-9] &2.47e-10 & [0,5$\times 10^{-16}$] & 2.87e-17\\
% \hline
% 5-segment & (2,1) & [0.018,0.030] &  0.026& [-5e-8,5e-8] &3.55e-10 & [0,2e-15] & 3.16$\times 10^{-16}$\\
% \hline  %在第一行和第二行之间绘制横线
% Prior range &[0.015,0.025]& [0,7e-10] & [0,8$\times 10^{-18}$]  \\
% \hline
\end{tabular}
\caption{The prior distributions for the three phenomenological parameters $\omega_0$, $\dot{\omega}_0$, and $\ddot{\omega}_0$ in the different time segments.}
\label{table:fittingparameters}
\end{table*}

\subsubsection{Results of the phenomenological waveform search}

%__run mcmc for diff segment__
For this search, we employ nested sampling to obtain the posterior distribution for each harmonic mode and each time segment to extensively explore the high posterior regions~\cite{skilling2004nested,speagle2020dynesty}. In addition to estimating the posterior distribution, nested sampling methods also can calculate the evidence by integrating the prior within nested ``shells'' of a constant likelihood. In practice, for each nested sampling execution, we use approximately $500$ core hours before stop the sampling. 

%__all mcmc results
We execute a total of 11 nested sampling operations. 
The results of nested sampling for each harmonic mode and each time segment are summarized in Table~\ref{table:fittingresults}. The results show the median and the 1-$\sigma$ range of the posterior distribution for the phenomenological parameters. We see that the bias of $\omega_{0}$ is less than the bias of $\dot{\omega}_{0}$, and the bias of $\dot{\omega}_{0}$ is smaller than the bias of $\ddot{\omega}_{0}$. This can be understood as the lower order term in the Taylor expansion of the phase evolution can be better tracked. 
In the first two time segments, the bias precision for $\omega_0$ is $\sim 0.1\,\%$, and it can even reach $10^{-6}$. Furthermore, the bias precision for the parameter $\dot{\omega}_{0}$ also remains within 10\,\%. However, as the observation time segments get closer to the plunge, the measured bias of the phenomenological parameters increases because the orbital evolution of the signal becomes faster, leading to the failure of the phenomenological waveform reproducing the signal accurately.%, as shown in Fig.~\ref{fig:p_e_mismatch}.

\begin{figure*}[tpb] \centering \includegraphics[height=0.8\textwidth, width=0.8\textwidth]{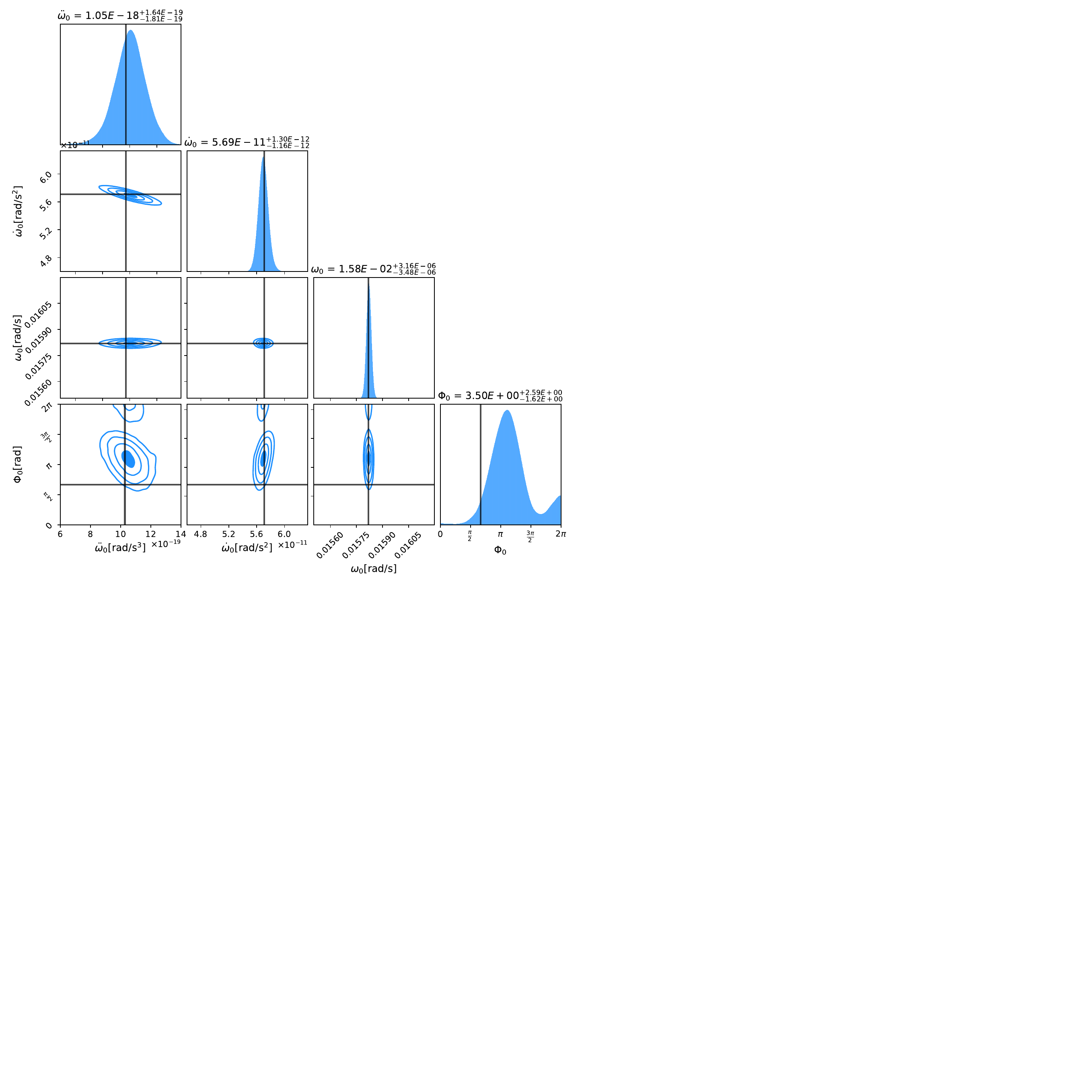}
    \caption{The posterior distribution of the four phenomenological parameters for the first time segment (2,0) mode.}
\label{fig:mcmcfit}
\end{figure*}

As an illustration, we display the results of the posterior distribution sampling for the phenomenological parameters of the $(2,0)$ harmonic mode for the first time segment in Fig.~\ref{fig:mcmcfit}. We see that the posterior distributions for the parameters $\omega$ and $\dot{\omega}$ exhibit very narrow and sharp peaks. The 1-$\sigma$ ranges for $\omega$ and $\dot{\omega}$ are 0.01\,\% and 1\,\% of the prior range, respectively while the 1-$\sigma$ range for $\ddot{\omega}$ can extend to 10\,\% of the prior range. However, for extrinsic parameters like the initial phase $\Phi_{0}$ and the positional parameters ($\theta,\phi,\psi$), the measurements are not accurate. This is because we do not take into account the evolution of the amplitude over time and the mutual coupling between these extrinsic parameters, causing the measurement results to deviate from the true values.

%%%--------------------------------------------------------------------------------%%%
%%% Revised on September 21 (end)

% \begin{table*}[htbp!]

\begin{sidewaystable*}[ht]
\centering
\resizebox{\textwidth}{40mm}{
\setlength{\tabcolsep}{1mm}{
\begin{tabular}{|c|c|c|c|c|c|c|c|c|c|c|c|}
 % & \multicolumn{3}{c|}{(2,2,0)}& \multicolumn{3}{c|}{(2,2,1)}& \multicolumn{3}{c|}{(2,2,2)}\\
% \hline  % 在表格最上方绘制横线
\hline 
\multirow{6}*{1-segment} & mode(m,n) & \multicolumn{3}{|c|}{(2,0)} & \multicolumn{3}{|c|}{(2,1)} &  \multicolumn{3}{|c|}{(2,2)} \\
\cline{2-11}
& parameters& $\omega_{0}/10^{-2} [rad/s]$ & $\dot{\omega}_{0}/10^{-11} [rad/s^2]$ & $\ddot{\omega}_{0}/10^{-18} [rad/s^3]$ &
$\omega_{0}/10^{-2} [rad/s]$ & $\dot{\omega}_{0}/10^{-11} [rad/s^2]$ & $\ddot{\omega}_{0}/10^{-18} [rad/s^3]$ & $\omega_{0}/10^{-2} [rad/s]$ & $\dot{\omega}_{0}/10^{-11} [rad/s^2]$ & $\ddot{\omega}_{0}/10^{-18} [rad/s^3]$ \\
\cline{2-11}
&signal&1.58&	5.71&	1.03&	2.01&	6.13&	0.97&	2.44&	6.55&	0.91 \\
\cline{2-11}
&measurement&$1.58_{-3.48 E-4}^{+3.16 E-4}$ &	$5.69_{-1.16E-1}^{+1.30E-1}$ & $1.05_{-1.81E-1}^{+1.64E-1}$  & $2.01 _{-2.24 E-4}^{+2.15 E-4}$&	$6.15_{-7.70E-2}^{+8.05E-2}$
& $0.95_{-1.10E-1}^{+1.07E-1}$&
$2.43 _{-2.57 E-4}^{+2.68 E-4}$&$6.43_{-9.56E-2}^{+9.23E-2}$
& $1.09_{-1.25E-1}^{+1.31E-1}$ \\
\cline{2-11}
& Bias& 0.00\% &	0.30 \% &	2.37\% &	0.00\% &	0.26\% &	1.76\% &	0.31\% &	1.82\% &	 19.72\% \\
\cline{2-11}
% & SNR & \multicolumn{3}{|c|}{6.39} & \multicolumn{3}{|c|}{11.19} & \multicolumn{3}{|c|}{7.85} \\ 
\hline 
\multirow{6}*{2-segment} & mode(m,n) & \multicolumn{3}{|c|}{(2,0)} & \multicolumn{3}{|c|}{(2,1)} &  \multicolumn{3}{|c|}{(2,2)} \\
\cline{2-11}
& parameters& $\frac{\omega_{0}}{10^{-2}}$ & $\frac{\dot{\omega}_{0}}{10^{-10}}$ & $\frac{\ddot{\omega}_{0}}{10^{-18}}$& $\frac{\omega_{0}}{10^{-2}}$ & $\frac{\dot{\omega}_{0}}{10^{-10}}$ & $\frac{\ddot{\omega}_{0}}{10^{-18}}$& $\frac{\omega_{0}}{10^{-2}}$ & $\frac{\dot{\omega}_{0}}{10^{-10}}$ & $\frac{\ddot{\omega}_{0}}{10^{-18}}$ \\
\cline{2-11}
&signal&1.66 &7.11 &1.73 &2.10 &7.46 &1.56 &2.53 &7.80 &1.39  \\
\cline{2-11}
&measurement& $1.66 _{-8.08 E-3}^{+3.68 E-4}$ & $7.10_{-1.35E-1}^{+1.12E-1}$
&$1.72_{-1.57E-1}^{+1.85E-1}$ & $2.10 _{-3.05 E-4}^{+1.27 E-3}$
&$6.86_{-3.11E-1}^{+1.09E-1}$&
$2.18_{-1.46E-1}^{+3.24E-1}$&$2.54_{-1.84 E-4}^{+1.97 E-4}$&
$7.50_{-7.94E-2}^{+7.32E-2}$
&$1.59_{-1.06E-1}^{+1.12E-1}$\\
% 1.67& 	1.34& 	6.58& 	2.10& 	1.50& 	4.42& 	2.43& 	1.24& 	4.87\\
\cline{2-11}
&Bias&0.00\%& 0.13\%&6.54\%&0.11\% & 7.99\%&39.80\%& 0.48 \%&3.96\%&14.5\%\\
\cline{2-11}
% & SNR & \multicolumn{3}{|c|}{8.23 } & \multicolumn{3}{|c|}{8.59 } & \multicolumn{3}{|c|}{5.89 } \\ 
\hline 
\multirow{6}*{3-segment} & mode(m,n) & \multicolumn{3}{|c|}{(2,0)} & \multicolumn{3}{|c|}{(2,1)} &  \multicolumn{3}{|c|}{(2,2)} \\
\cline{2-11}
& parameters& $\frac{\omega_{0}}{10^{-2}}$ & $\frac{\dot{\omega}_{0}}{10^{-10}}$ & $\frac{\ddot{\omega}_{0}}{10^{-18}}$& $\frac{\omega_{0}}{10^{-2}}$ & $\frac{\dot{\omega}_{0}}{10^{-10}}$ & $\frac{\ddot{\omega}_{0}}{10^{-18}}$& $\frac{\omega_{0}}{10^{-2}}$ & $\frac{\dot{\omega}_{0}}{10^{-10}}$ & $\frac{\ddot{\omega}_{0}}{10^{-18}}$ \\
\cline{2-11}
&signal&1.77& 9.32 & 3.57 &2.20& 9.49 & 3.01 & 2.64 & 9.66 &2.46  \\
\cline{2-11}
&measurement&
$1.68_{-6.17 E-2}^{+1.03 E-1}$&$7.54_{-11.7}^{+2.22}$&$26.4_-21.2^{+36.5}$ &
$1.94_{-1.44 E-1}^{+3.64 E-1}$ &$3.24_{-7.76}^{+5.29}$&$17.5_{-15.2}^{+48.0}$&
$2.40_{-9.56E-2}^{+2.37 E-1}$&$1.81_{-6.04}^{+6.51}$&$19.6_{-13.6}^{+34.3}$\\
% 1.56& 	0.40& 	138.23& 	2.24& 	0.31& 	29.95& 	2.49& 	0.64& 	154.98  \\
\cline{2-11}
&Bias& 5.09\% & 19.17\% & 637.71\% &	11.74\%&  65.84\% & 480.89\%&	9.05\%&	87.33\%& 699.28\% \\
\cline{2-11}
% & SNR & \multicolumn{3}{|c|}{5.44} & \multicolumn{3}{|c|}{5.53} & \multicolumn{3}{|c|}{5.57 } \\ 
\hline
\end{tabular}
}}
\caption{The measurement results of phenomenological parameters ($\omega_{0},\dot{\omega}_{0},\ddot{\omega}_{0}$) with different harmonics for different time segments.}
\label{table:fittingresults}
% \end{table*}
\end{sidewaystable*}

%%% Revised on September 22 (start)
%%%--------------------------------------------------------------------------------%%%

\subsection{Hierarchical search for physical parameters}
\label{sec:inversion}

Achieving the data analysis of EMRI signals without extensive prior knowledge of the physical parameters poses a formidable challenge since expanding the parameter space for the search entails a commensurate increase in computational demands. For example, in previous works the parameter range for $\log_{10}(M/{\rm M_\odot})$ was within $\sim0.05$ for a harmonic search~\cite{babak2009algorithm}, it was within $\sim0.2$ when using a MCMC method~\cite{chua2022nonlocal}, and for a phenomenological search $M$ was fixed~\cite{Wang2012xh}. In this study, we combined various search methods to perform a hierarchical search for the physical parameters considering a parameter range for $M$ over two orders of magnitudes. 
%\ato{What range did you use for your search? It would be good to highlight how much you can expand the parameter range.} 
%\ato{Also this part is nice but quite general. Maybe we should move it to the conclusions.}
%\Yecq{Here is check the parameter range of differ method, along with links:} http://172.16.209.129/yechq6/check\_parameter\_range\_for\_different\_method.\Yecq{It's worth noting that the unit of M here is a base-10 logarithm (lg) and not a natural logarithm (log) with the base 'e'.}

Our ultimate goal is to obtain the range for the physical parameters using a hierarchical search method based on the highly precise measurement results of the phenomenological parameters in Sec.~\ref{sec:analysisphenom} and the rough range of physical parameters obtained in Sec.~\ref{sec:harmonicsearch}. The phenomenological parameters of different harmonics at various time segments provide strong constraints on the physical parameters and thus we attempt to map the posterior distribution of the phenomenological parameters to the physical parameters.

%___introduce parameter inference
\subsubsection{Search method}
\label{sec:inversion:method}

It is difficult to directly convert from phenomenological waveform parameters to physical parameters. On the one hand, there is no analytical formula to deduce the physical parameters directly from the phenomenological parameters. On the other hand, the multipeak structure of the posterior distribution of physical parameters means that it might be one-to-many mapping from phenomenological parameters.

Here, we employ a grid-based approach to the physical parameters to facilitate the resolution of this procedure. For each point on the grid, we use FEW to calculate the harmonic phase evolution for each time segment. Then, we fit them using Eq.~(\ref{eq:phase_fitting}) to obtain the corresponding phenomenological parameters. In the Schwarzschild eccentricity condition, the parameters $[M,\mu,p_{0},e_{0}]$ directly determine the trajectory of the phase evolution.
Therefore, the set of four physical parameters can be transformed into the phenomenological parameter space. Here, we choose 18 phenomenological parameters for the first two time segments with three main harmonic modes in each segment and the three parameters ($\omega_{0},\dot{\omega}_{0},\ddot{\omega}_{0}$) for each harmonic mode. This choice is based on the exceptional accuracy of the parameter measurement in these segments when compared to the other time segments (see Table~\ref{table:fittingresults}). Moreover, we further choose a set of nine phenomenological parameters, which include the $(2,0)$ and $(2,1)$ modes for the first time segment and the $(2,0)$ mode for the second time segment, which are the closest to the signal in terms of their corresponding phenomenological parameters and will be used for a finer differentiation in the search.

For each phenomenological parameter, we check if the parameter falls within the 3-$\sigma$ posterior distribution obtained from the nested sampling. In principle, one can depict the physical parameters using a sufficiently dense grid. 
The number of such ``close" parameters is used to rank the fitness.
However, to account for practical computational constraints we first employ a coarse grid to the region of the parameter space obtained from the harmonic mode search and then a finer grid to a smaller region identified during the search with the coarse grid.

\subsubsection{Results of the hierarchical search}

\begin{figure*}[tpbh] \centering 
\includegraphics[width=1\textwidth]{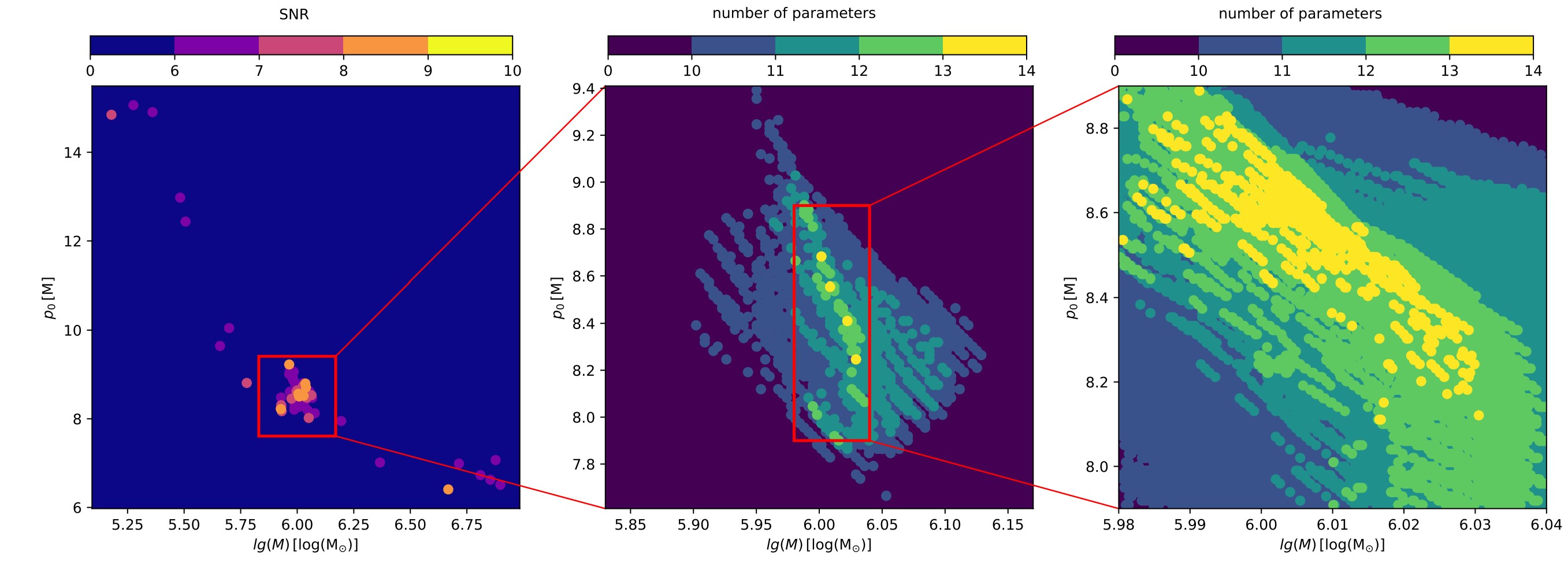}
    \caption{The results of the hierarchical search for physical parameters. The left graph shows the search result using the physical harmonic waveforms. The middle graph presents the search results from the coarse grid and the results from the fine grid search are shown in the right graph. Different search methods and strategies were employed here, as detailed in Sec.~\ref{sec:harmonicsearch} and \ref{sec:inversion}.}
\label{fig:hierarchical_search}
\end{figure*}

Figure ~\ref{fig:hierarchical_search} shows the hierarchical research of physical parameters, showcasing the outcomes at each stage of the search process. The left image represents the search results obtained from the physical harmonic search. The middle and right images depict the outcomes obtained through coarse and fine grid-based methods, respectively.

From the distribution of parameters from the harmonic template search introduced in Section~\ref{sec:harmonicsearch}, we obtain the parameter range ($[5.85,6.17]$ for $\log_{10}(M/{\rm M_\odot})$ and $[7.68,9.48]\,M$ for $p_0$) shown in the red box in the left graph of Fig.~\ref{fig:hierarchical_search}. For the grid-based search, the parameter range is determined based on the number of parameters for which the phenomenological parameters fall within the 3-$\sigma$ posterior distribution obtained in Sec.~\ref{sec:analysisphenom}. This is not equivalent to the criteria in the first step.

%, as here, two entirely different methods are employed to obtain the parameter ranges. In particular, we use that for the grid-based search we conducted high-precision measurements of the phenomenological parameters. Therefore, in the grid-based search, there is no need to perform calculations of the waveforms and the likelihood functions; we simply need to transform the physical parameters into phenomenological parameters.

\begin{figure*}[tpbh] \centering \includegraphics[height=0.8\textwidth]{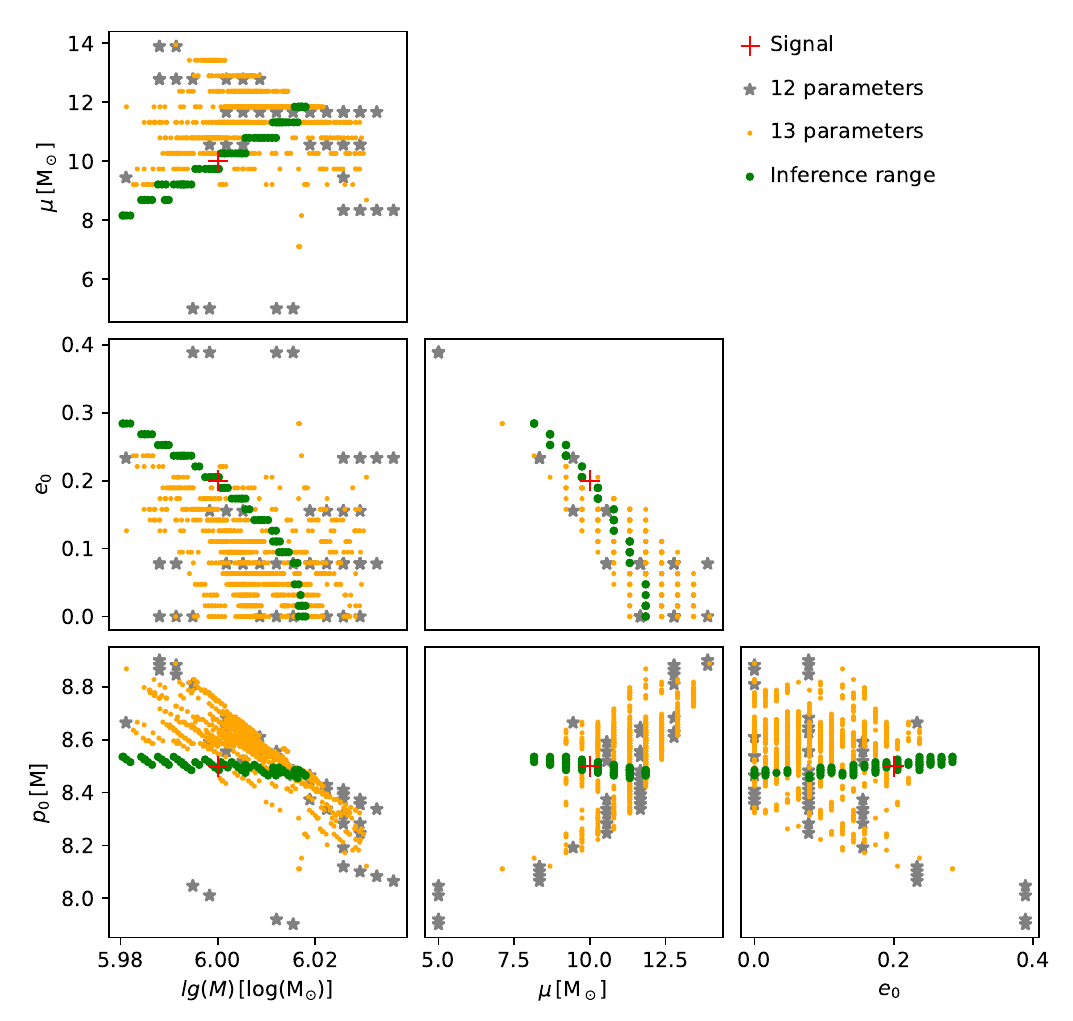}
    \caption{The hierarchical search results for the four physical parameters obtained from the coarse grid and the fine grid. The terms `12 parameters' and `13 parameters' refer to the presence of 12 phenomenological parameters within the 3-$\sigma$ range of the coarse grid and 13 phenomenological parameters within the 3-$\sigma$ range of the fine grid, respectively. The inference range indicates that a specific set of nine parameters that were measured with particularly high precision falls within the 3-$\sigma$ range.}
\label{fig:inversion}
\end{figure*}

For the coarse grid, we uniformly sample 100 points for the parameters $M$ and $p_0$ within the range of the first step results. Additionally, for the parameters $e_0$ and $\mu$, we uniformly sample 10 points within the parameter ranges $[0,0.7]$ and $[5,15]{\mathrm M_\odot}$, respectively. This grid comprises a total of 1,000,000 points and the overall computational cost amounts to approximately 500 core hours. In this grid, 94 best-fit grid points, each contains 12 phenomenological parameters that fall within the 3-$\sigma$ range, are shown in the center graph of Fig.~\ref{fig:hierarchical_search}. These phenomenological parameters correspond to the range of physical parameters $\log_{10}(M/{\mathrm M_\odot}) \in [5.98,6.04]$, $p_{0} \in [7.9,8.9]M$, $e_{0} \in [0,0.4]$, and $\mu \in [8,13]{\mathrm M_\odot}$ and are shown as gray stars in Fig.~\ref{fig:inversion}.

For the fine grid, the grid parameters' range is based on the results of the coarse grid. Similarly to the previous search, we uniformly sample 100 points for the parameters $M$ and $p_0$, and 10 points for $e_0$ but 50 points for $\mu$. For each parameter point, we find the best fit phenomenological parameters to its physical waveforms in all 3 segments, and count how many of these parameters fall within the $3-\sigma$ range of the recovered parameters. 
Among all grid points, 807 of them (yellow points in Fig. ~\ref{fig:inversion}) have 13 close enough phenomenological parameters, however, those parameters that do not match could be significantly biased.
Upon closer examination, we determined if we assign different weights among parameters, or more specifically, if we focus on the nine high-accuracy parameters discussed in Section ~\ref{sec:inversion:method}, we can obtain better fitting.
We then add an additional requirement, that all nine high-accuracy parameters have to be located within the $3-\sigma$ region.
With this new condition, we show in Fig. ~\ref{fig:inversion} the points that contain 12 or 13 parameters located close enough to the recovered parameters, nine of them being the high-accuracy parameters. 
We can observe that this best-fit region is narrow, and it contains the injected physical parameter.

This is due to the presence of different harmonics and segmented phase evolution information, which inherently imposes strong constraints on the physical parameters themselves. In the end, we constrain the range of physical parameters to
$[5.980,6.018]$ for $\log_{10}(M/{\rm M_\odot})$, $[8.455,8.535]\,M$ for $p_0$, $[0,0.284]$ for $e_0$, and $[8.158,11.842]\,{\rm M_\odot}$ for $\mu$.

Based on high-precision measurements of the phenomenological parameters and the physical parameter ranges obtained from the search results of the physical waveforms, we achieve the relative error in the measurement of $M$ to be within 4\,\% while the smallest errors for the other parameters can be within 0.5\,\%. This parameter accuracy is sufficient for the requirement of future EMRI parameter inference~\cite{chua2021non}.

\section{conclusions and discusion}\label{sec:conclusion}

%=== Conclusion ===

Accomplishing the identification of EMRI signals and providing ranges for the physical parameters for subsequent parameter inference is challenging. On the one hand, performing searches using only physical waveforms encounters difficulties due to the multipeak structure of the posterior distribution of physical parameters which makes it difficult for stochastic sampling algorithms to navigate between peaks. Even employing something like the ``one-step" likelihood function approach does not provide global guidance for the region of the high posterior regions. 
On the other hand, when only employing phenomenological waveforms for the searches, the lack of prior information about phenomenological and physical parameters makes signal searching and the inference of physical parameters difficult. Therefore, we adopt in this work a combined approach of both methods. Phenomenological waveform searches effectively avoid the multipeak structure of the posterior distribution of physical parameters, while physical parameter searches provide prior information to narrow down the search space of parameters.

We have achieved for the first time the identification of EMRI signals without any additional prior information on physical parameters. High-precision measurements of EMRI signals have been achieved using a hierarchical search that combines the search for physical parameters that guides subsequent parameter inference and a semicoherent search with phenomenological waveforms that reaches precision levels down to $10^{-4}$ for the phenomenological waveform parameters $\omega_{0}$, $\dot{\omega}_{0}$, and $\ddot{\omega}_{0}$. As a result, we obtain measurement relative errors of less than 4\,\% for the mass of the MBH, while keeping the relative errors of the other parameters within as small as 0.5\,\%.

Although we only search for an EMRI signal assuming a Schwarzschild eccentric background, we believe the method presented can be used universally for the search of EMRI signals for several reasons:
\begin{itemize}
    \item [1)]
    In Ref.~\cite{babak2009} the applicability of the harmonic detection in AK waveforms under the Kerr eccentric background was demonstrated. Once a sufficient number of harmonic templates match the signal, a physical parameter range can be established. Our objective in this step is not only to obtain the range of physical parameters but also to precisely extract the phase evolution information. Furthermore, we employ the more accurate fully relativistic FEW waveforms to obtain a more realistic phase evolution.
    
    \item [2)]
    In Ref.~\cite{wang2012extreme}, the amplitude evolution was also neglected, and an expansion of the phase as a Taylor series of third order was also used. Our goal in this step was not only to achieve the matching for phase of the signal but, more importantly, to achieve high-precision measurements of phenomenological parameters which is crucial for the hierarchical search of the physical parameters.
    
    \item [3)] 
    The difference between the Schwarzschild background and the Kerr background induces a relatively slow difference in the frequency evolution in the early inspiral phase of the EMRI signal. In Fig.\ref{fig:p_e_mismatch}, the mismatch between the phenomenological and physical waveforms implies that there can be a high level of correlation during the early inspiral phase of the EMRI signal. Therefore, within a short time segment, the signal can be well-matched using the phenomenological waveform. In addition, FEW interpolates cubic splines of sparse phases to obtain complete phase evolution~\cite{chua2021rapid,katz2021fast}. The farther away from the plunge in time, the sparser the phase evolution is calculated. Hence,  fitting the phase within shorter time segments at earlier stages yields improved fitting outcomes.
\end{itemize}
% \ato{How you are explaining 1) and 2), sounds like the idea and work were done before by the others and you are just repeating it... Write it differently.}
% % 这句话这样写有问题，这里想说的是把施瓦西黑洞波形拓展到kerr 黑洞背景同样地适用。
% 3) Whether it is the Schwarzschild eccentricity background or the Kerr black hole background, in the early inspiral phase of the EMRI signal, its frequency evolution is relatively slow.
% Therefore, within a short time segment, the signal can be well matched using the phenomenological waveform.

%____future___

For the example event, $\sim 45$ hours, $\sim 40$ hours, and $\sim 15$ hours are consumed used on a 100 core cluster for the harmonic mode search stage, the semi-coherent search stage, and the hierarchical search stage.
We remind the readers that we did not consider the time delay interference(TDI) response in the calculation, however, including the first-generation TDI should not induce a significant computational burden, since the generation speed of the latest frequency-domain waveforms is faster than time-domain waveforms, making the use of a full TDI response more computationally efficient~\cite{Speri2023FastAF}. However, the second-generation TDI response which can not be easily computed in the frequency domain would be more challenging. 
In the future, we wish to extend the analysis to realistic cases where multiple EMRI events might exist simultaneously. In the simplest case, we can perform the search-and-remove to identify EMRI signals one by one. In this case, the computational cost would be linear to the event number.
A more desirable pipeline, known as the global analysis method,  which can simultaneously process multiple signals might be more complicated and computationally more demanding ~\cite{littenberg2023prototype,cornish2021bayeswave}.
However, the advance of searching algorithms together with the upgrading of the computing hardware make it possible to solve the EMRI identification problem in the future.

For this proof-of-principle study, we implement the analysis in different stages separately, and human intervention is needed at the end of each stage, to extract the parameter range and to set the initial conditions in the following stages. Once the process is fixed, it is possible to automate the whole process and no human intervention would be mandatory.

% \Yecq{Even with the full TDI response employed for future searches, computation time will not significantly increase.The generation speed of the latest frequency-domain waveforms is faster than time-domain waveforms, making the use of a full TDI response more computationally efficient~\cite{Speri2023FastAF}. Moreover,in future observational data, the simultaneous presence of multiple EMRI signals may lead to confusion arising from the overlap between these EMRIs. 
% One feasible approach is to conduct a pipeline global analysis of multiple source types~\cite{littenberg2023prototype,cornish2021bayeswave}.
% Specifically, this entails initially extracting and isolating the strongest signal from the data, followed by the analysis of the subsequent signal, and so forth.
% Currently, the analysis of a single source takes $\sim 1$ week, with about $20\%$ of this time dedicated to human analysis of each step's results, which are then used as initial conditions for the next step of the search. In the future, setting more reasonable thresholds can enable automated analysis of the results, thereby achieving automation of the entire process.Finally,}

% \sout{Moreover,} the high-precision measurements based on phenomenological parameters could also be used for the spatial localization of EMRI events. The effectivity of converting the phenomenological waveform parameters into physical parameters in this case should be studied in more detail.

In the future, more complex scenarios to approach the final EMRI signal processing pipeline should be studied. For example, the application of fully relativistic Kerr waveforms for the identification should be explored as well as investigating the more intricate TDI response.Moreover, the high-precision measurements based on phenomenological parameters could also be used for the spatial localization of EMRI events. The effectivity of converting the phenomenological waveform parameters into physical parameters in this case should be studied in more detail.To better achieve these objectives and reproduce this work, the entire pipeline can be found at: \href{https://github.com/ChangQingYe-SYSU/EMRI_identification}{https://github.com/ChangQingYe\-SYSU/EMRI\_identification.}

% Additionally, the unimodal structure of the posterior distribution of phenomenological waveform parameters guides us in overcoming the challenges posed by the multimodal structure of physical parameters in EMRI signal analysis.

%%%--------------------------------------------------------------------------------%%%
%%% Revised on September 22 (end)

\section*{Acknowledgments}

The authors thank En-Kun Li, Jianwei Mei, Zheng Wu, Han Wang, and Shuo Sun for helpful discussions.  
This work has been supported by Guangdong Major Project of Basic and Applied Basic Research (Grant No. 2019B030302001), the Natural Science Foundation of China (Grants No. 12173104 and No. 12261131504), Hebei Natural Science Foundation with Grant No. A2023201041,Guangdong Basic and Applied Basic Research Foundation(Grant No. 2023A1515030116), ATO acknowledges support from the China Postdoctoral Science Foundation (Grant No. 2022M723676). 
The authors acknowledge the uses of the calculating utilities of \textsf{NUMPY}
\cite{vanderWalt:2011bqk}, \textsf{SCIPY} \cite{Virtanen:2019joe}, and the
plotting utilities of \textsf{MATPLOTLIB} \cite{Hunter:2007ouj}.

\bibliography{ref}

\begin{thebibliography}{74}%
\makeatletter
\providecommand \@ifxundefined [1]{%
 \@ifx{#1\undefined}
}%
\providecommand \@ifnum [1]{%
 \ifnum #1\expandafter \@firstoftwo
 \else \expandafter \@secondoftwo
 \fi
}%
\providecommand \@ifx [1]{%
 \ifx #1\expandafter \@firstoftwo
 \else \expandafter \@secondoftwo
 \fi
}%
\providecommand \natexlab [1]{#1}%
\providecommand \enquote  [1]{``#1''}%
\providecommand \bibnamefont  [1]{#1}%
\providecommand \bibfnamefont [1]{#1}%
\providecommand \citenamefont [1]{#1}%
\providecommand \href@noop [0]{\@secondoftwo}%
\providecommand \href [0]{\begingroup \@sanitize@url \@href}%
\providecommand \@href[1]{\@@startlink{#1}\@@href}%
\providecommand \@@href[1]{\endgroup#1\@@endlink}%
\providecommand \@sanitize@url [0]{\catcode `\\12\catcode `\$12\catcode
  `\&12\catcode `\#12\catcode `\^12\catcode `\_12\catcode `\%12\relax}%
\providecommand \@@startlink[1]{}%
\providecommand \@@endlink[0]{}%
\providecommand \url  [0]{\begingroup\@sanitize@url \@url }%
\providecommand \@url [1]{\endgroup\@href {#1}{\urlprefix }}%
\providecommand \urlprefix  [0]{URL }%
\providecommand \Eprint [0]{\href }%
\providecommand \doibase [0]{http://dx.doi.org/}%
\providecommand \selectlanguage [0]{\@gobble}%
\providecommand \bibinfo  [0]{\@secondoftwo}%
\providecommand \bibfield  [0]{\@secondoftwo}%
\providecommand \translation [1]{[#1]}%
\providecommand \BibitemOpen [0]{}%
\providecommand \bibitemStop [0]{}%
\providecommand \bibitemNoStop [0]{.\EOS\space}%
\providecommand \EOS [0]{\spacefactor3000\relax}%
\providecommand \BibitemShut  [1]{\csname bibitem#1\endcsname}%
\let\auto@bib@innerbib\@empty
%</preamble>
\bibitem [{\citenamefont {Abbott}\ \emph {et~al.}(2019)\citenamefont {Abbott}
  \emph {et~al.}}]{LIGOScientific:2018mvr}%
  \BibitemOpen
  \bibfield  {author} {\bibinfo {author} {\bibfnamefont {B.~P.}\ \bibnamefont
  {Abbott}} \emph {et~al.} (\bibinfo {collaboration} {LIGO Scientific,
  Virgo}),\ }\href {\doibase 10.1103/PhysRevX.9.031040} {\bibfield  {journal}
  {\bibinfo  {journal} {Phys. Rev. X}\ }\textbf {\bibinfo {volume} {9}},\
  \bibinfo {pages} {031040} (\bibinfo {year} {2019})},\ \Eprint
  {http://arxiv.org/abs/1811.12907} {arXiv:1811.12907 [astro-ph.HE]}
  \BibitemShut {NoStop}%
\bibitem [{\citenamefont {Abbott}\ \emph {et~al.}(2021)\citenamefont {Abbott},
  \citenamefont {Abbott}, \citenamefont {Abraham}, \citenamefont {Acernese},
  \citenamefont {Ackley}, \citenamefont {Adams}, \citenamefont {Adams},
  \citenamefont {Adhikari}, \citenamefont {Adya}, \citenamefont {Affeldt} \emph
  {et~al.}}]{abbott2021gwtc}%
  \BibitemOpen
  \bibfield  {author} {\bibinfo {author} {\bibfnamefont {R.}~\bibnamefont
  {Abbott}}, \bibinfo {author} {\bibfnamefont {T.}~\bibnamefont {Abbott}},
  \bibinfo {author} {\bibfnamefont {S.}~\bibnamefont {Abraham}}, \bibinfo
  {author} {\bibfnamefont {F.}~\bibnamefont {Acernese}}, \bibinfo {author}
  {\bibfnamefont {K.}~\bibnamefont {Ackley}}, \bibinfo {author} {\bibfnamefont
  {A.}~\bibnamefont {Adams}}, \bibinfo {author} {\bibfnamefont
  {C.}~\bibnamefont {Adams}}, \bibinfo {author} {\bibfnamefont
  {R.}~\bibnamefont {Adhikari}}, \bibinfo {author} {\bibfnamefont
  {V.}~\bibnamefont {Adya}}, \bibinfo {author} {\bibfnamefont {C.}~\bibnamefont
  {Affeldt}},  \emph {et~al.},\ }\href@noop {} {\bibfield  {journal} {\bibinfo
  {journal} {Physical Review X}\ }\textbf {\bibinfo {volume} {11}},\ \bibinfo
  {pages} {021053} (\bibinfo {year} {2021})}\BibitemShut {NoStop}%
\bibitem [{\citenamefont {Abbott}\ \emph {et~al.}(2023)\citenamefont {Abbott},
  \citenamefont {Abbott}, \citenamefont {Acernese}, \citenamefont {Ackley},
  \citenamefont {Adams}, \citenamefont {Adhikari}, \citenamefont {Adhikari},
  \citenamefont {Adya}, \citenamefont {Affeldt}, \citenamefont {Agarwal} \emph
  {et~al.}}]{abbott2023gwtc}%
  \BibitemOpen
  \bibfield  {author} {\bibinfo {author} {\bibfnamefont {R.}~\bibnamefont
  {Abbott}}, \bibinfo {author} {\bibfnamefont {T.}~\bibnamefont {Abbott}},
  \bibinfo {author} {\bibfnamefont {F.}~\bibnamefont {Acernese}}, \bibinfo
  {author} {\bibfnamefont {K.}~\bibnamefont {Ackley}}, \bibinfo {author}
  {\bibfnamefont {C.}~\bibnamefont {Adams}}, \bibinfo {author} {\bibfnamefont
  {N.}~\bibnamefont {Adhikari}}, \bibinfo {author} {\bibfnamefont
  {R.}~\bibnamefont {Adhikari}}, \bibinfo {author} {\bibfnamefont
  {V.}~\bibnamefont {Adya}}, \bibinfo {author} {\bibfnamefont {C.}~\bibnamefont
  {Affeldt}}, \bibinfo {author} {\bibfnamefont {D.}~\bibnamefont {Agarwal}},
  \emph {et~al.},\ }\href@noop {} {\bibfield  {journal} {\bibinfo  {journal}
  {Physical Review X}\ }\textbf {\bibinfo {volume} {13}},\ \bibinfo {pages}
  {041039} (\bibinfo {year} {2023})}\BibitemShut {NoStop}%
\bibitem [{\citenamefont {Danzmann}\ and\ \citenamefont
  {R{\"u}diger}(2003)}]{danzmann2003lisa}%
  \BibitemOpen
  \bibfield  {author} {\bibinfo {author} {\bibfnamefont {K.}~\bibnamefont
  {Danzmann}}\ and\ \bibinfo {author} {\bibfnamefont {A.}~\bibnamefont
  {R{\"u}diger}},\ }\href@noop {} {\bibfield  {journal} {\bibinfo  {journal}
  {Classical and Quantum Gravity}\ }\textbf {\bibinfo {volume} {20}},\ \bibinfo
  {pages} {S1} (\bibinfo {year} {2003})}\BibitemShut {NoStop}%
\bibitem [{\citenamefont {Amaro-Seoane}\ \emph {et~al.}(2013)\citenamefont
  {Amaro-Seoane} \emph {et~al.}}]{AmaroSeoane:2012km}%
  \BibitemOpen
  \bibfield  {author} {\bibinfo {author} {\bibfnamefont {P.}~\bibnamefont
  {Amaro-Seoane}} \emph {et~al.},\ }\href@noop {} {\bibfield  {journal}
  {\bibinfo  {journal} {GW Notes}\ }\textbf {\bibinfo {volume} {6}},\ \bibinfo
  {pages} {4} (\bibinfo {year} {2013})},\ \Eprint
  {http://arxiv.org/abs/1201.3621} {arXiv:1201.3621 [astro-ph.CO]} \BibitemShut
  {NoStop}%
\bibitem [{\citenamefont {Amaro-Seoane}\ \emph {et~al.}(2017)\citenamefont
  {Amaro-Seoane}, \citenamefont {Audley}, \citenamefont {Babak}, \citenamefont
  {Baker}, \citenamefont {Barausse}, \citenamefont {Bender}, \citenamefont
  {Berti}, \citenamefont {Binetruy}, \citenamefont {Born}, \citenamefont
  {Bortoluzzi} \emph {et~al.}}]{amaro2017laser}%
  \BibitemOpen
  \bibfield  {author} {\bibinfo {author} {\bibfnamefont {P.}~\bibnamefont
  {Amaro-Seoane}}, \bibinfo {author} {\bibfnamefont {H.}~\bibnamefont
  {Audley}}, \bibinfo {author} {\bibfnamefont {S.}~\bibnamefont {Babak}},
  \bibinfo {author} {\bibfnamefont {J.}~\bibnamefont {Baker}}, \bibinfo
  {author} {\bibfnamefont {E.}~\bibnamefont {Barausse}}, \bibinfo {author}
  {\bibfnamefont {P.}~\bibnamefont {Bender}}, \bibinfo {author} {\bibfnamefont
  {E.}~\bibnamefont {Berti}}, \bibinfo {author} {\bibfnamefont
  {P.}~\bibnamefont {Binetruy}}, \bibinfo {author} {\bibfnamefont
  {M.}~\bibnamefont {Born}}, \bibinfo {author} {\bibfnamefont {D.}~\bibnamefont
  {Bortoluzzi}},  \emph {et~al.},\ }\href@noop {} {\bibfield  {journal}
  {\bibinfo  {journal} {arXiv preprint arXiv:1702.00786}\ } (\bibinfo {year}
  {2017})}\BibitemShut {NoStop}%
\bibitem [{\citenamefont {Mei}\ \emph {et~al.}(2020)\citenamefont {Mei},
  \citenamefont {Bai}, \citenamefont {Bao}, \citenamefont {Barausse},
  \citenamefont {Cai}, \citenamefont {Canuto}, \citenamefont {Cao},
  \citenamefont {Chen}, \citenamefont {Chen}, \citenamefont {Ding} \emph
  {et~al.}}]{mei2020tianqin}%
  \BibitemOpen
  \bibfield  {author} {\bibinfo {author} {\bibfnamefont {J.}~\bibnamefont
  {Mei}}, \bibinfo {author} {\bibfnamefont {Y.-Z.}\ \bibnamefont {Bai}},
  \bibinfo {author} {\bibfnamefont {J.}~\bibnamefont {Bao}}, \bibinfo {author}
  {\bibfnamefont {E.}~\bibnamefont {Barausse}}, \bibinfo {author}
  {\bibfnamefont {L.}~\bibnamefont {Cai}}, \bibinfo {author} {\bibfnamefont
  {E.}~\bibnamefont {Canuto}}, \bibinfo {author} {\bibfnamefont
  {B.}~\bibnamefont {Cao}}, \bibinfo {author} {\bibfnamefont {W.-M.}\
  \bibnamefont {Chen}}, \bibinfo {author} {\bibfnamefont {Y.}~\bibnamefont
  {Chen}}, \bibinfo {author} {\bibfnamefont {Y.-W.}\ \bibnamefont {Ding}},
  \emph {et~al.},\ }\href@noop {} {\bibfield  {journal} {\bibinfo  {journal}
  {Progress of Theoretical and Experimental Physics}\ } (\bibinfo {year}
  {2020})}\BibitemShut {NoStop}%
\bibitem [{\citenamefont {{Torres-Orjuela}}\ \emph {et~al.}(2023)\citenamefont
  {{Torres-Orjuela}}, \citenamefont {{Huang}}, \citenamefont {{Liang}},
  \citenamefont {{Liu}}, \citenamefont {{Wang}}, \citenamefont {{Ye}},
  \citenamefont {{Hu}},\ and\ \citenamefont
  {{Mei}}}]{torres-orjuela_huang_2023}%
  \BibitemOpen
  \bibfield  {author} {\bibinfo {author} {\bibfnamefont {A.}~\bibnamefont
  {{Torres-Orjuela}}}, \bibinfo {author} {\bibfnamefont {S.-J.}\ \bibnamefont
  {{Huang}}}, \bibinfo {author} {\bibfnamefont {Z.-C.}\ \bibnamefont
  {{Liang}}}, \bibinfo {author} {\bibfnamefont {S.}~\bibnamefont {{Liu}}},
  \bibinfo {author} {\bibfnamefont {H.-T.}\ \bibnamefont {{Wang}}}, \bibinfo
  {author} {\bibfnamefont {C.-Q.}\ \bibnamefont {{Ye}}}, \bibinfo {author}
  {\bibfnamefont {Y.-M.}\ \bibnamefont {{Hu}}}, \ and\ \bibinfo {author}
  {\bibfnamefont {J.}~\bibnamefont {{Mei}}},\ }\href {\doibase
  10.48550/arXiv.2307.16628} {\bibfield  {journal} {\bibinfo  {journal} {arXiv
  e-prints}\ ,\ \bibinfo {eid} {arXiv:2307.16628}} (\bibinfo {year} {2023})},\
  \Eprint {http://arxiv.org/abs/2307.16628} {arXiv:2307.16628 [gr-qc]}
  \BibitemShut {NoStop}%
\bibitem [{\citenamefont {Amaro-Seoane}\ \emph {et~al.}(2007)\citenamefont
  {Amaro-Seoane}, \citenamefont {Gair}, \citenamefont {Freitag}, \citenamefont
  {Coleman~Miller}, \citenamefont {Mandel}, \citenamefont {Cutler},\ and\
  \citenamefont {Babak}}]{AmaroSeoane:2007aw}%
  \BibitemOpen
  \bibfield  {author} {\bibinfo {author} {\bibfnamefont {P.}~\bibnamefont
  {Amaro-Seoane}}, \bibinfo {author} {\bibfnamefont {J.~R.}\ \bibnamefont
  {Gair}}, \bibinfo {author} {\bibfnamefont {M.}~\bibnamefont {Freitag}},
  \bibinfo {author} {\bibfnamefont {M.}~\bibnamefont {Coleman~Miller}},
  \bibinfo {author} {\bibfnamefont {I.}~\bibnamefont {Mandel}}, \bibinfo
  {author} {\bibfnamefont {C.~J.}\ \bibnamefont {Cutler}}, \ and\ \bibinfo
  {author} {\bibfnamefont {S.}~\bibnamefont {Babak}},\ }\href {\doibase
  10.1088/0264-9381/24/17/R01} {\bibfield  {journal} {\bibinfo  {journal}
  {Class. Quant. Grav.}\ }\textbf {\bibinfo {volume} {24}},\ \bibinfo {pages}
  {R113} (\bibinfo {year} {2007})},\ \Eprint
  {http://arxiv.org/abs/astro-ph/0703495} {arXiv:astro-ph/0703495} \BibitemShut
  {NoStop}%
\bibitem [{\citenamefont {Amaro-Seoane}(2018)}]{AmaroSeoane:2012tx}%
  \BibitemOpen
  \bibfield  {author} {\bibinfo {author} {\bibfnamefont {P.}~\bibnamefont
  {Amaro-Seoane}},\ }\href {\doibase 10.1007/s41114-018-0013-8} {\bibfield
  {journal} {\bibinfo  {journal} {Living Rev. Rel.}\ }\textbf {\bibinfo
  {volume} {21}},\ \bibinfo {pages} {4} (\bibinfo {year} {2018})},\ \Eprint
  {http://arxiv.org/abs/1205.5240} {arXiv:1205.5240 [astro-ph.CO]} \BibitemShut
  {NoStop}%
\bibitem [{\citenamefont {Babak}\ \emph {et~al.}(2017)\citenamefont {Babak},
  \citenamefont {Gair}, \citenamefont {Sesana}, \citenamefont {Barausse},
  \citenamefont {Sopuerta}, \citenamefont {Berry}, \citenamefont {Berti},
  \citenamefont {Amaro-Seoane}, \citenamefont {Petiteau},\ and\ \citenamefont
  {Klein}}]{babak2017science}%
  \BibitemOpen
  \bibfield  {author} {\bibinfo {author} {\bibfnamefont {S.}~\bibnamefont
  {Babak}}, \bibinfo {author} {\bibfnamefont {J.}~\bibnamefont {Gair}},
  \bibinfo {author} {\bibfnamefont {A.}~\bibnamefont {Sesana}}, \bibinfo
  {author} {\bibfnamefont {E.}~\bibnamefont {Barausse}}, \bibinfo {author}
  {\bibfnamefont {C.~F.}\ \bibnamefont {Sopuerta}}, \bibinfo {author}
  {\bibfnamefont {C.~P.}\ \bibnamefont {Berry}}, \bibinfo {author}
  {\bibfnamefont {E.}~\bibnamefont {Berti}}, \bibinfo {author} {\bibfnamefont
  {P.}~\bibnamefont {Amaro-Seoane}}, \bibinfo {author} {\bibfnamefont
  {A.}~\bibnamefont {Petiteau}}, \ and\ \bibinfo {author} {\bibfnamefont
  {A.}~\bibnamefont {Klein}},\ }\href@noop {} {\bibfield  {journal} {\bibinfo
  {journal} {Physical Review D}\ }\textbf {\bibinfo {volume} {95}},\ \bibinfo
  {pages} {103012} (\bibinfo {year} {2017})}\BibitemShut {NoStop}%
\bibitem [{\citenamefont {Gair}\ \emph {et~al.}(2017)\citenamefont {Gair},
  \citenamefont {Babak}, \citenamefont {Sesana}, \citenamefont {Amaro-Seoane},
  \citenamefont {Barausse}, \citenamefont {Berry}, \citenamefont {Berti},\ and\
  \citenamefont {Sopuerta}}]{gair2017prospects}%
  \BibitemOpen
  \bibfield  {author} {\bibinfo {author} {\bibfnamefont {J.~R.}\ \bibnamefont
  {Gair}}, \bibinfo {author} {\bibfnamefont {S.}~\bibnamefont {Babak}},
  \bibinfo {author} {\bibfnamefont {A.}~\bibnamefont {Sesana}}, \bibinfo
  {author} {\bibfnamefont {P.}~\bibnamefont {Amaro-Seoane}}, \bibinfo {author}
  {\bibfnamefont {E.}~\bibnamefont {Barausse}}, \bibinfo {author}
  {\bibfnamefont {C.~P.}\ \bibnamefont {Berry}}, \bibinfo {author}
  {\bibfnamefont {E.}~\bibnamefont {Berti}}, \ and\ \bibinfo {author}
  {\bibfnamefont {C.}~\bibnamefont {Sopuerta}},\ }in\ \href@noop {} {\emph
  {\bibinfo {booktitle} {Journal of Physics: Conference Series}}},\ Vol.\
  \bibinfo {volume} {840}\ (\bibinfo {organization} {IOP Publishing},\ \bibinfo
  {year} {2017})\ p.\ \bibinfo {pages} {012021}\BibitemShut {NoStop}%
\bibitem [{\citenamefont {Fan}\ \emph {et~al.}(2020)\citenamefont {Fan},
  \citenamefont {Hu}, \citenamefont {Barausse}, \citenamefont {Sesana},
  \citenamefont {Zhang}, \citenamefont {Zhang}, \citenamefont {Zi},
  \citenamefont {Mei} \emph {et~al.}}]{fan2020science}%
  \BibitemOpen
  \bibfield  {author} {\bibinfo {author} {\bibfnamefont {H.-M.}\ \bibnamefont
  {Fan}}, \bibinfo {author} {\bibfnamefont {Y.-M.}\ \bibnamefont {Hu}},
  \bibinfo {author} {\bibfnamefont {E.}~\bibnamefont {Barausse}}, \bibinfo
  {author} {\bibfnamefont {A.}~\bibnamefont {Sesana}}, \bibinfo {author}
  {\bibfnamefont {J.-d.}\ \bibnamefont {Zhang}}, \bibinfo {author}
  {\bibfnamefont {X.}~\bibnamefont {Zhang}}, \bibinfo {author} {\bibfnamefont
  {T.-G.}\ \bibnamefont {Zi}}, \bibinfo {author} {\bibfnamefont
  {J.}~\bibnamefont {Mei}},  \emph {et~al.},\ }\href@noop {} {\bibfield
  {journal} {\bibinfo  {journal} {Physical Review D}\ }\textbf {\bibinfo
  {volume} {102}},\ \bibinfo {pages} {063016} (\bibinfo {year}
  {2020})}\BibitemShut {NoStop}%
\bibitem [{\citenamefont {Zi}\ \emph {et~al.}(2021)\citenamefont {Zi},
  \citenamefont {Zhang}, \citenamefont {Fan}, \citenamefont {Zhang},
  \citenamefont {Hu}, \citenamefont {Shi},\ and\ \citenamefont
  {Mei}}]{Zi:2021pdp}%
  \BibitemOpen
  \bibfield  {author} {\bibinfo {author} {\bibfnamefont {T.-G.}\ \bibnamefont
  {Zi}}, \bibinfo {author} {\bibfnamefont {J.-D.}\ \bibnamefont {Zhang}},
  \bibinfo {author} {\bibfnamefont {H.-M.}\ \bibnamefont {Fan}}, \bibinfo
  {author} {\bibfnamefont {X.-T.}\ \bibnamefont {Zhang}}, \bibinfo {author}
  {\bibfnamefont {Y.-M.}\ \bibnamefont {Hu}}, \bibinfo {author} {\bibfnamefont
  {C.}~\bibnamefont {Shi}}, \ and\ \bibinfo {author} {\bibfnamefont
  {J.}~\bibnamefont {Mei}},\ }\href {\doibase 10.1103/PhysRevD.104.064008}
  {\bibfield  {journal} {\bibinfo  {journal} {Phys. Rev. D}\ }\textbf {\bibinfo
  {volume} {104}},\ \bibinfo {pages} {064008} (\bibinfo {year} {2021})},\
  \Eprint {http://arxiv.org/abs/2104.06047} {arXiv:2104.06047 [gr-qc]}
  \BibitemShut {NoStop}%
\bibitem [{\citenamefont {Maselli}\ \emph {et~al.}(2022)\citenamefont
  {Maselli}, \citenamefont {Franchini}, \citenamefont {Gualtieri},
  \citenamefont {Sotiriou}, \citenamefont {Barsanti},\ and\ \citenamefont
  {Pani}}]{Maselli:2021men}%
  \BibitemOpen
  \bibfield  {author} {\bibinfo {author} {\bibfnamefont {A.}~\bibnamefont
  {Maselli}}, \bibinfo {author} {\bibfnamefont {N.}~\bibnamefont {Franchini}},
  \bibinfo {author} {\bibfnamefont {L.}~\bibnamefont {Gualtieri}}, \bibinfo
  {author} {\bibfnamefont {T.~P.}\ \bibnamefont {Sotiriou}}, \bibinfo {author}
  {\bibfnamefont {S.}~\bibnamefont {Barsanti}}, \ and\ \bibinfo {author}
  {\bibfnamefont {P.}~\bibnamefont {Pani}},\ }\href {\doibase
  10.1038/s41550-021-01589-5} {\bibfield  {journal} {\bibinfo  {journal}
  {Nature Astron.}\ }\textbf {\bibinfo {volume} {6}},\ \bibinfo {pages} {464}
  (\bibinfo {year} {2022})},\ \Eprint {http://arxiv.org/abs/2106.11325}
  {arXiv:2106.11325 [gr-qc]} \BibitemShut {NoStop}%
\bibitem [{\citenamefont {Barsanti}\ \emph {et~al.}(2022)\citenamefont
  {Barsanti}, \citenamefont {Franchini}, \citenamefont {Gualtieri},
  \citenamefont {Maselli},\ and\ \citenamefont {Sotiriou}}]{Barsanti:2022ana}%
  \BibitemOpen
  \bibfield  {author} {\bibinfo {author} {\bibfnamefont {S.}~\bibnamefont
  {Barsanti}}, \bibinfo {author} {\bibfnamefont {N.}~\bibnamefont {Franchini}},
  \bibinfo {author} {\bibfnamefont {L.}~\bibnamefont {Gualtieri}}, \bibinfo
  {author} {\bibfnamefont {A.}~\bibnamefont {Maselli}}, \ and\ \bibinfo
  {author} {\bibfnamefont {T.~P.}\ \bibnamefont {Sotiriou}},\ }\href@noop {}
  {\enquote {\bibinfo {title} {{Extreme mass-ratio inspirals as probes of
  scalar fields: eccentric equatorial orbits around Kerr black holes}},}\ }
  (\bibinfo {year} {2022}),\ \Eprint {http://arxiv.org/abs/2203.05003}
  {arXiv:2203.05003 [gr-qc]} \BibitemShut {NoStop}%
\bibitem [{\citenamefont {Torres-Orjuela}\ \emph {et~al.}(2021)\citenamefont
  {Torres-Orjuela}, \citenamefont {Seoane}, \citenamefont {Xuan}, \citenamefont
  {Chua}, \citenamefont {Rosell},\ and\ \citenamefont
  {Chen}}]{torres2021exciting}%
  \BibitemOpen
  \bibfield  {author} {\bibinfo {author} {\bibfnamefont {A.}~\bibnamefont
  {Torres-Orjuela}}, \bibinfo {author} {\bibfnamefont {P.~A.}\ \bibnamefont
  {Seoane}}, \bibinfo {author} {\bibfnamefont {Z.}~\bibnamefont {Xuan}},
  \bibinfo {author} {\bibfnamefont {A.~J.}\ \bibnamefont {Chua}}, \bibinfo
  {author} {\bibfnamefont {M.~J.}\ \bibnamefont {Rosell}}, \ and\ \bibinfo
  {author} {\bibfnamefont {X.}~\bibnamefont {Chen}},\ }\href@noop {} {\bibfield
   {journal} {\bibinfo  {journal} {Physical Review Letters}\ }\textbf {\bibinfo
  {volume} {127}},\ \bibinfo {pages} {041102} (\bibinfo {year}
  {2021})}\BibitemShut {NoStop}%
\bibitem [{\citenamefont {{Barausse}}\ and\ \citenamefont
  {{Rezzolla}}(2008)}]{barausse_rezzolla_2008}%
  \BibitemOpen
  \bibfield  {author} {\bibinfo {author} {\bibfnamefont {E.}~\bibnamefont
  {{Barausse}}}\ and\ \bibinfo {author} {\bibfnamefont {L.}~\bibnamefont
  {{Rezzolla}}},\ }\href {\doibase 10.1103/PhysRevD.77.104027} {\bibfield
  {journal} {\bibinfo  {journal} {\prd}\ }\textbf {\bibinfo {volume} {77}},\
  \bibinfo {eid} {104027} (\bibinfo {year} {2008})},\ \Eprint
  {http://arxiv.org/abs/0711.4558} {arXiv:0711.4558 [gr-qc]} \BibitemShut
  {NoStop}%
\bibitem [{\citenamefont {Pan}\ \emph {et~al.}(2021)\citenamefont {Pan},
  \citenamefont {Lyu},\ and\ \citenamefont {Yang}}]{Pan:2021oob}%
  \BibitemOpen
  \bibfield  {author} {\bibinfo {author} {\bibfnamefont {Z.}~\bibnamefont
  {Pan}}, \bibinfo {author} {\bibfnamefont {Z.}~\bibnamefont {Lyu}}, \ and\
  \bibinfo {author} {\bibfnamefont {H.}~\bibnamefont {Yang}},\ }\href {\doibase
  10.1103/PhysRevD.104.063007} {\bibfield  {journal} {\bibinfo  {journal}
  {Phys. Rev. D}\ }\textbf {\bibinfo {volume} {104}},\ \bibinfo {pages}
  {063007} (\bibinfo {year} {2021})},\ \Eprint
  {http://arxiv.org/abs/2104.01208} {arXiv:2104.01208 [astro-ph.HE]}
  \BibitemShut {NoStop}%
\bibitem [{\citenamefont {Fan}\ \emph {et~al.}(2022)\citenamefont {Fan},
  \citenamefont {Zhong}, \citenamefont {Liang}, \citenamefont {Wu},
  \citenamefont {Zhang},\ and\ \citenamefont {Hu}}]{Fan:2022wio}%
  \BibitemOpen
  \bibfield  {author} {\bibinfo {author} {\bibfnamefont {H.-M.}\ \bibnamefont
  {Fan}}, \bibinfo {author} {\bibfnamefont {S.}~\bibnamefont {Zhong}}, \bibinfo
  {author} {\bibfnamefont {Z.-C.}\ \bibnamefont {Liang}}, \bibinfo {author}
  {\bibfnamefont {Z.}~\bibnamefont {Wu}}, \bibinfo {author} {\bibfnamefont
  {J.-d.}\ \bibnamefont {Zhang}}, \ and\ \bibinfo {author} {\bibfnamefont
  {Y.-M.}\ \bibnamefont {Hu}},\ }\href {\doibase 10.1103/PhysRevD.106.124028}
  {\bibfield  {journal} {\bibinfo  {journal} {Phys. Rev. D}\ }\textbf {\bibinfo
  {volume} {106}},\ \bibinfo {pages} {124028} (\bibinfo {year} {2022})},\
  \Eprint {http://arxiv.org/abs/2209.13387} {arXiv:2209.13387 [gr-qc]}
  \BibitemShut {NoStop}%
\bibitem [{\citenamefont {Yunes}\ \emph {et~al.}(2011)\citenamefont {Yunes},
  \citenamefont {Kocsis}, \citenamefont {Loeb},\ and\ \citenamefont
  {Haiman}}]{Yunes2011}%
  \BibitemOpen
  \bibfield  {author} {\bibinfo {author} {\bibfnamefont {N.}~\bibnamefont
  {Yunes}}, \bibinfo {author} {\bibfnamefont {B.}~\bibnamefont {Kocsis}},
  \bibinfo {author} {\bibfnamefont {A.}~\bibnamefont {Loeb}}, \ and\ \bibinfo
  {author} {\bibfnamefont {Z.}~\bibnamefont {Haiman}},\ }\href {\doibase
  10.1103/PhysRevLett.107.171103} {\bibfield  {journal} {\bibinfo  {journal}
  {\prl}\ }\textbf {\bibinfo {volume} {107}},\ \bibinfo {pages} {171103}
  (\bibinfo {year} {2011})}\BibitemShut {NoStop}%
\bibitem [{\citenamefont {Barausse}\ \emph {et~al.}(2015)\citenamefont
  {Barausse}, \citenamefont {Cardoso},\ and\ \citenamefont
  {Pani}}]{Barausse2015}%
  \BibitemOpen
  \bibfield  {author} {\bibinfo {author} {\bibfnamefont {E.}~\bibnamefont
  {Barausse}}, \bibinfo {author} {\bibfnamefont {V.}~\bibnamefont {Cardoso}}, \
  and\ \bibinfo {author} {\bibfnamefont {P.}~\bibnamefont {Pani}},\ }in\ \href
  {\doibase 10.1088/1742-6596/610/1/012044} {\emph {\bibinfo {booktitle}
  {Journal of Physics Conference Series}}},\ \bibinfo {series} {Journal of
  Physics Conference Series}, Vol.\ \bibinfo {volume} {610}\ (\bibinfo {year}
  {2015})\ p.\ \bibinfo {pages} {12044}\BibitemShut {NoStop}%
\bibitem [{\citenamefont {Cardoso}\ \emph {et~al.}(2022)\citenamefont
  {Cardoso}, \citenamefont {Destounis}, \citenamefont {Duque}, \citenamefont
  {Macedo},\ and\ \citenamefont {Maselli}}]{cardoso2022gravitational}%
  \BibitemOpen
  \bibfield  {author} {\bibinfo {author} {\bibfnamefont {V.}~\bibnamefont
  {Cardoso}}, \bibinfo {author} {\bibfnamefont {K.}~\bibnamefont {Destounis}},
  \bibinfo {author} {\bibfnamefont {F.}~\bibnamefont {Duque}}, \bibinfo
  {author} {\bibfnamefont {R.~P.}\ \bibnamefont {Macedo}}, \ and\ \bibinfo
  {author} {\bibfnamefont {A.}~\bibnamefont {Maselli}},\ }\href@noop {}
  {\bibfield  {journal} {\bibinfo  {journal} {Physical Review Letters}\
  }\textbf {\bibinfo {volume} {129}},\ \bibinfo {pages} {241103} (\bibinfo
  {year} {2022})}\BibitemShut {NoStop}%
\bibitem [{\citenamefont {Gair}\ \emph {et~al.}(2010)\citenamefont {Gair},
  \citenamefont {Tang},\ and\ \citenamefont {Volonteri}}]{Gair2010}%
  \BibitemOpen
  \bibfield  {author} {\bibinfo {author} {\bibfnamefont {J.~R.}\ \bibnamefont
  {Gair}}, \bibinfo {author} {\bibfnamefont {C.}~\bibnamefont {Tang}}, \ and\
  \bibinfo {author} {\bibfnamefont {M.}~\bibnamefont {Volonteri}},\ }\href
  {\doibase 10.1103/PhysRevD.81.104014} {\bibfield  {journal} {\bibinfo
  {journal} {\prd}\ }\textbf {\bibinfo {volume} {81}},\ \bibinfo {pages}
  {104014} (\bibinfo {year} {2010})}\BibitemShut {NoStop}%
\bibitem [{\citenamefont {MacLeod}\ and\ \citenamefont
  {Hogan}(2008)}]{MacLeod:2007jd}%
  \BibitemOpen
  \bibfield  {author} {\bibinfo {author} {\bibfnamefont {C.~L.}\ \bibnamefont
  {MacLeod}}\ and\ \bibinfo {author} {\bibfnamefont {C.~J.}\ \bibnamefont
  {Hogan}},\ }\href {\doibase 10.1103/PhysRevD.77.043512} {\bibfield  {journal}
  {\bibinfo  {journal} {Phys. Rev. D}\ }\textbf {\bibinfo {volume} {77}},\
  \bibinfo {pages} {043512} (\bibinfo {year} {2008})},\ \Eprint
  {http://arxiv.org/abs/0712.0618} {arXiv:0712.0618 [astro-ph]} \BibitemShut
  {NoStop}%
\bibitem [{\citenamefont {{Laghi}}\ \emph {et~al.}(2021)\citenamefont
  {{Laghi}}, \citenamefont {{Tamanini}}, \citenamefont {{Del Pozzo}},
  \citenamefont {{Sesana}}, \citenamefont {{Gair}},\ and\ \citenamefont
  {{Babak}}}]{Laghi2021EMRICosmology}%
  \BibitemOpen
  \bibfield  {author} {\bibinfo {author} {\bibfnamefont {D.}~\bibnamefont
  {{Laghi}}}, \bibinfo {author} {\bibfnamefont {N.}~\bibnamefont {{Tamanini}}},
  \bibinfo {author} {\bibfnamefont {W.}~\bibnamefont {{Del Pozzo}}}, \bibinfo
  {author} {\bibfnamefont {A.}~\bibnamefont {{Sesana}}}, \bibinfo {author}
  {\bibfnamefont {J.}~\bibnamefont {{Gair}}}, \ and\ \bibinfo {author}
  {\bibfnamefont {S.}~\bibnamefont {{Babak}}},\ }\href@noop {} {\bibfield
  {journal} {\bibinfo  {journal} {arXiv e-prints}\ ,\ \bibinfo {eid}
  {arXiv:2102.01708}} (\bibinfo {year} {2021})},\ \Eprint
  {http://arxiv.org/abs/2102.01708} {arXiv:2102.01708 [astro-ph.CO]}
  \BibitemShut {NoStop}%
\bibitem [{\citenamefont {Hughes}\ \emph {et~al.}(2005)\citenamefont {Hughes},
  \citenamefont {Drasco}, \citenamefont {Flanagan},\ and\ \citenamefont
  {Franklin}}]{hughes2005gravitational}%
  \BibitemOpen
  \bibfield  {author} {\bibinfo {author} {\bibfnamefont {S.~A.}\ \bibnamefont
  {Hughes}}, \bibinfo {author} {\bibfnamefont {S.}~\bibnamefont {Drasco}},
  \bibinfo {author} {\bibfnamefont {E.~E.}\ \bibnamefont {Flanagan}}, \ and\
  \bibinfo {author} {\bibfnamefont {J.}~\bibnamefont {Franklin}},\ }\href@noop
  {} {\bibfield  {journal} {\bibinfo  {journal} {Physical review letters}\
  }\textbf {\bibinfo {volume} {94}},\ \bibinfo {pages} {221101} (\bibinfo
  {year} {2005})}\BibitemShut {NoStop}%
\bibitem [{\citenamefont {van~de Meent}(2018)}]{self-force_Van}%
  \BibitemOpen
  \bibfield  {author} {\bibinfo {author} {\bibfnamefont {M.}~\bibnamefont
  {van~de Meent}},\ }\href {\doibase 10.1103/PhysRevD.97.104033} {\bibfield
  {journal} {\bibinfo  {journal} {Phys. Rev. D}\ }\textbf {\bibinfo {volume}
  {97}},\ \bibinfo {pages} {104033} (\bibinfo {year} {2018})}\BibitemShut
  {NoStop}%
\bibitem [{\citenamefont {Barack}(2009)}]{PT_Barack}%
  \BibitemOpen
  \bibfield  {author} {\bibinfo {author} {\bibfnamefont {L.}~\bibnamefont
  {Barack}},\ }\href {\doibase 10.1088/0264-9381/26/21/213001} {\bibfield
  {journal} {\bibinfo  {journal} {Class. Quant. Grav.}\ }\textbf {\bibinfo
  {volume} {26}},\ \bibinfo {pages} {213001} (\bibinfo {year} {2009})},\
  \Eprint {http://arxiv.org/abs/0908.1664} {arXiv:0908.1664 [gr-qc]}
  \BibitemShut {NoStop}%
\bibitem [{\citenamefont {{Barack}}\ and\ \citenamefont
  {{Cutler}}(2004)}]{AK_2004_PRD}%
  \BibitemOpen
  \bibfield  {author} {\bibinfo {author} {\bibfnamefont {L.}~\bibnamefont
  {{Barack}}}\ and\ \bibinfo {author} {\bibfnamefont {C.}~\bibnamefont
  {{Cutler}}},\ }\href {\doibase 10.1103/PhysRevD.69.082005} {\bibfield
  {journal} {\bibinfo  {journal} {\prd}\ }\textbf {\bibinfo {volume} {69}},\
  \bibinfo {eid} {082005} (\bibinfo {year} {2004})},\ \Eprint
  {http://arxiv.org/abs/gr-qc/0310125} {arXiv:gr-qc/0310125 [gr-qc]}
  \BibitemShut {NoStop}%
\bibitem [{\citenamefont {Babak}\ \emph {et~al.}(2007)\citenamefont {Babak},
  \citenamefont {Fang}, \citenamefont {Gair}, \citenamefont {Glampedakis},\
  and\ \citenamefont {Hughes}}]{NK_2007_PRD}%
  \BibitemOpen
  \bibfield  {author} {\bibinfo {author} {\bibfnamefont {S.}~\bibnamefont
  {Babak}}, \bibinfo {author} {\bibfnamefont {H.}~\bibnamefont {Fang}},
  \bibinfo {author} {\bibfnamefont {J.~R.}\ \bibnamefont {Gair}}, \bibinfo
  {author} {\bibfnamefont {K.}~\bibnamefont {Glampedakis}}, \ and\ \bibinfo
  {author} {\bibfnamefont {S.~A.}\ \bibnamefont {Hughes}},\ }\href {\doibase
  10.1103/PhysRevD.75.024005} {\bibfield  {journal} {\bibinfo  {journal} {Phys.
  Rev. D}\ }\textbf {\bibinfo {volume} {75}},\ \bibinfo {pages} {024005}
  (\bibinfo {year} {2007})}\BibitemShut {NoStop}%
\bibitem [{\citenamefont {Chua}\ \emph {et~al.}(2017)\citenamefont {Chua},
  \citenamefont {Moore},\ and\ \citenamefont {Gair}}]{chua2017augmented}%
  \BibitemOpen
  \bibfield  {author} {\bibinfo {author} {\bibfnamefont {A.~J.}\ \bibnamefont
  {Chua}}, \bibinfo {author} {\bibfnamefont {C.~J.}\ \bibnamefont {Moore}}, \
  and\ \bibinfo {author} {\bibfnamefont {J.~R.}\ \bibnamefont {Gair}},\
  }\href@noop {} {\bibfield  {journal} {\bibinfo  {journal} {Physical Review
  D}\ }\textbf {\bibinfo {volume} {96}},\ \bibinfo {pages} {044005} (\bibinfo
  {year} {2017})}\BibitemShut {NoStop}%
\bibitem [{\citenamefont {Chua}\ \emph {et~al.}(2021)\citenamefont {Chua},
  \citenamefont {Katz}, \citenamefont {Warburton},\ and\ \citenamefont
  {Hughes}}]{chua2021rapid}%
  \BibitemOpen
  \bibfield  {author} {\bibinfo {author} {\bibfnamefont {A.~J.}\ \bibnamefont
  {Chua}}, \bibinfo {author} {\bibfnamefont {M.~L.}\ \bibnamefont {Katz}},
  \bibinfo {author} {\bibfnamefont {N.}~\bibnamefont {Warburton}}, \ and\
  \bibinfo {author} {\bibfnamefont {S.~A.}\ \bibnamefont {Hughes}},\
  }\href@noop {} {\bibfield  {journal} {\bibinfo  {journal} {Physical Review
  Letters}\ }\textbf {\bibinfo {volume} {126}},\ \bibinfo {pages} {051102}
  (\bibinfo {year} {2021})}\BibitemShut {NoStop}%
\bibitem [{\citenamefont {Katz}\ \emph {et~al.}(2021)\citenamefont {Katz},
  \citenamefont {Chua}, \citenamefont {Speri}, \citenamefont {Warburton},\ and\
  \citenamefont {Hughes}}]{katz2021fast}%
  \BibitemOpen
  \bibfield  {author} {\bibinfo {author} {\bibfnamefont {M.~L.}\ \bibnamefont
  {Katz}}, \bibinfo {author} {\bibfnamefont {A.~J.}\ \bibnamefont {Chua}},
  \bibinfo {author} {\bibfnamefont {L.}~\bibnamefont {Speri}}, \bibinfo
  {author} {\bibfnamefont {N.}~\bibnamefont {Warburton}}, \ and\ \bibinfo
  {author} {\bibfnamefont {S.~A.}\ \bibnamefont {Hughes}},\ }\href@noop {}
  {\bibfield  {journal} {\bibinfo  {journal} {Physical Review D}\ }\textbf
  {\bibinfo {volume} {104}},\ \bibinfo {pages} {064047} (\bibinfo {year}
  {2021})}\BibitemShut {NoStop}%
\bibitem [{\citenamefont {Chua}\ and\ \citenamefont
  {Cutler}(2022)}]{chua2022nonlocal}%
  \BibitemOpen
  \bibfield  {author} {\bibinfo {author} {\bibfnamefont {A.~J.}\ \bibnamefont
  {Chua}}\ and\ \bibinfo {author} {\bibfnamefont {C.~J.}\ \bibnamefont
  {Cutler}},\ }\href@noop {} {\bibfield  {journal} {\bibinfo  {journal}
  {Physical Review D}\ }\textbf {\bibinfo {volume} {106}},\ \bibinfo {pages}
  {124046} (\bibinfo {year} {2022})}\BibitemShut {NoStop}%
\bibitem [{\citenamefont {Gair}\ \emph {et~al.}(2004)\citenamefont {Gair},
  \citenamefont {Barack}, \citenamefont {Creighton}, \citenamefont {Cutler},
  \citenamefont {Larson}, \citenamefont {Phinney},\ and\ \citenamefont
  {Vallisneri}}]{gair2004event}%
  \BibitemOpen
  \bibfield  {author} {\bibinfo {author} {\bibfnamefont {J.~R.}\ \bibnamefont
  {Gair}}, \bibinfo {author} {\bibfnamefont {L.}~\bibnamefont {Barack}},
  \bibinfo {author} {\bibfnamefont {T.}~\bibnamefont {Creighton}}, \bibinfo
  {author} {\bibfnamefont {C.}~\bibnamefont {Cutler}}, \bibinfo {author}
  {\bibfnamefont {S.~L.}\ \bibnamefont {Larson}}, \bibinfo {author}
  {\bibfnamefont {E.~S.}\ \bibnamefont {Phinney}}, \ and\ \bibinfo {author}
  {\bibfnamefont {M.}~\bibnamefont {Vallisneri}},\ }\href@noop {} {\bibfield
  {journal} {\bibinfo  {journal} {Classical and Quantum Gravity}\ }\textbf
  {\bibinfo {volume} {21}},\ \bibinfo {pages} {S1595} (\bibinfo {year}
  {2004})}\BibitemShut {NoStop}%
\bibitem [{\citenamefont {Chua}(2022)}]{chua2022one}%
  \BibitemOpen
  \bibfield  {author} {\bibinfo {author} {\bibfnamefont {A.~J.}\ \bibnamefont
  {Chua}},\ }\href@noop {} {\bibfield  {journal} {\bibinfo  {journal} {Physical
  Review D}\ }\textbf {\bibinfo {volume} {106}},\ \bibinfo {pages} {104051}
  (\bibinfo {year} {2022})}\BibitemShut {NoStop}%
\bibitem [{\citenamefont {Zhang}\ \emph {et~al.}(2022)\citenamefont {Zhang},
  \citenamefont {Messenger}, \citenamefont {Korsakova}, \citenamefont {Chan},
  \citenamefont {Hu},\ and\ \citenamefont {Zhang}}]{zhang2022detecting}%
  \BibitemOpen
  \bibfield  {author} {\bibinfo {author} {\bibfnamefont {X.-T.}\ \bibnamefont
  {Zhang}}, \bibinfo {author} {\bibfnamefont {C.}~\bibnamefont {Messenger}},
  \bibinfo {author} {\bibfnamefont {N.}~\bibnamefont {Korsakova}}, \bibinfo
  {author} {\bibfnamefont {M.~L.}\ \bibnamefont {Chan}}, \bibinfo {author}
  {\bibfnamefont {Y.-M.}\ \bibnamefont {Hu}}, \ and\ \bibinfo {author}
  {\bibfnamefont {J.-d.}\ \bibnamefont {Zhang}},\ }\href@noop {} {\bibfield
  {journal} {\bibinfo  {journal} {Physical Review D}\ }\textbf {\bibinfo
  {volume} {105}},\ \bibinfo {pages} {123027} (\bibinfo {year}
  {2022})}\BibitemShut {NoStop}%
\bibitem [{\citenamefont {Wang}\ \emph
  {et~al.}(2012{\natexlab{a}})\citenamefont {Wang}, \citenamefont {Shang},
  \citenamefont {Babak}, \citenamefont {Shang},\ and\ \citenamefont
  {Babak}}]{Wang2012xh}%
  \BibitemOpen
  \bibfield  {author} {\bibinfo {author} {\bibfnamefont {Y.}~\bibnamefont
  {Wang}}, \bibinfo {author} {\bibfnamefont {Y.}~\bibnamefont {Shang}},
  \bibinfo {author} {\bibfnamefont {S.}~\bibnamefont {Babak}}, \bibinfo
  {author} {\bibfnamefont {Y.}~\bibnamefont {Shang}}, \ and\ \bibinfo {author}
  {\bibfnamefont {S.}~\bibnamefont {Babak}},\ }\href {\doibase
  10.1103/PhysRevD.86.104050} {\bibfield  {journal} {\bibinfo  {journal} {Phys.
  Rev. D}\ }\textbf {\bibinfo {volume} {86}},\ \bibinfo {pages} {104050}
  (\bibinfo {year} {2012}{\natexlab{a}})},\ \Eprint
  {http://arxiv.org/abs/1207.4956} {arXiv:1207.4956 [gr-qc]} \BibitemShut
  {NoStop}%
\bibitem [{\citenamefont {Babak}\ \emph
  {et~al.}(2009{\natexlab{a}})\citenamefont {Babak}, \citenamefont {Gair},\
  and\ \citenamefont {Porter}}]{babak2009algorithm}%
  \BibitemOpen
  \bibfield  {author} {\bibinfo {author} {\bibfnamefont {S.}~\bibnamefont
  {Babak}}, \bibinfo {author} {\bibfnamefont {J.~R.}\ \bibnamefont {Gair}}, \
  and\ \bibinfo {author} {\bibfnamefont {E.~K.}\ \bibnamefont {Porter}},\
  }\href@noop {} {\bibfield  {journal} {\bibinfo  {journal} {Classical and
  quantum gravity}\ }\textbf {\bibinfo {volume} {26}},\ \bibinfo {pages}
  {135004} (\bibinfo {year} {2009}{\natexlab{a}})}\BibitemShut {NoStop}%
\bibitem [{\citenamefont {Wang}\ \emph {et~al.}(2015)\citenamefont {Wang},
  \citenamefont {Heinzel},\ and\ \citenamefont {Danzmann}}]{wang2015first}%
  \BibitemOpen
  \bibfield  {author} {\bibinfo {author} {\bibfnamefont {Y.}~\bibnamefont
  {Wang}}, \bibinfo {author} {\bibfnamefont {G.}~\bibnamefont {Heinzel}}, \
  and\ \bibinfo {author} {\bibfnamefont {K.}~\bibnamefont {Danzmann}},\
  }\href@noop {} {\bibfield  {journal} {\bibinfo  {journal} {Physical Review
  D}\ }\textbf {\bibinfo {volume} {92}},\ \bibinfo {pages} {044037} (\bibinfo
  {year} {2015})}\BibitemShut {NoStop}%
\bibitem [{\citenamefont {Gair}\ and\ \citenamefont
  {Wen}(2005)}]{gair2005detecting}%
  \BibitemOpen
  \bibfield  {author} {\bibinfo {author} {\bibfnamefont {J.}~\bibnamefont
  {Gair}}\ and\ \bibinfo {author} {\bibfnamefont {L.}~\bibnamefont {Wen}},\
  }\href@noop {} {\bibfield  {journal} {\bibinfo  {journal} {Classical and
  Quantum Gravity}\ }\textbf {\bibinfo {volume} {22}},\ \bibinfo {pages}
  {S1359} (\bibinfo {year} {2005})}\BibitemShut {NoStop}%
\bibitem [{\citenamefont {Wen}\ and\ \citenamefont
  {Gair}(2005)}]{wen2005detecting}%
  \BibitemOpen
  \bibfield  {author} {\bibinfo {author} {\bibfnamefont {L.}~\bibnamefont
  {Wen}}\ and\ \bibinfo {author} {\bibfnamefont {J.~R.}\ \bibnamefont {Gair}},\
  }\href@noop {} {\bibfield  {journal} {\bibinfo  {journal} {Classical and
  Quantum Gravity}\ }\textbf {\bibinfo {volume} {22}},\ \bibinfo {pages} {S445}
  (\bibinfo {year} {2005})}\BibitemShut {NoStop}%
\bibitem [{\citenamefont {Gair}\ \emph
  {et~al.}(2008{\natexlab{a}})\citenamefont {Gair}, \citenamefont {Mandel},\
  and\ \citenamefont {Wen}}]{Gair:2008ec}%
  \BibitemOpen
  \bibfield  {author} {\bibinfo {author} {\bibfnamefont {J.~R.}\ \bibnamefont
  {Gair}}, \bibinfo {author} {\bibfnamefont {I.}~\bibnamefont {Mandel}}, \ and\
  \bibinfo {author} {\bibfnamefont {L.}~\bibnamefont {Wen}},\ }\href {\doibase
  10.1088/0264-9381/25/18/184031} {\bibfield  {journal} {\bibinfo  {journal}
  {Class. Quant. Grav.}\ }\textbf {\bibinfo {volume} {25}},\ \bibinfo {pages}
  {184031} (\bibinfo {year} {2008}{\natexlab{a}})},\ \Eprint
  {http://arxiv.org/abs/0804.1084} {arXiv:0804.1084 [gr-qc]} \BibitemShut
  {NoStop}%
\bibitem [{\citenamefont {Gair}\ \emph
  {et~al.}(2008{\natexlab{b}})\citenamefont {Gair}, \citenamefont {Porter},
  \citenamefont {Babak},\ and\ \citenamefont {Barack}}]{Gair2008a}%
  \BibitemOpen
  \bibfield  {author} {\bibinfo {author} {\bibfnamefont {J.~R.}\ \bibnamefont
  {Gair}}, \bibinfo {author} {\bibfnamefont {E.}~\bibnamefont {Porter}},
  \bibinfo {author} {\bibfnamefont {S.}~\bibnamefont {Babak}}, \ and\ \bibinfo
  {author} {\bibfnamefont {L.}~\bibnamefont {Barack}},\ }\href {\doibase
  10.1088/0264-9381/25/18/184030} {\bibfield  {journal} {\bibinfo  {journal}
  {Classical and Quantum Gravity}\ }\textbf {\bibinfo {volume} {25}},\ \bibinfo
  {pages} {184030} (\bibinfo {year} {2008}{\natexlab{b}})}\BibitemShut
  {NoStop}%
\bibitem [{\citenamefont {Chua}(2016)}]{Chua:2016jnd}%
  \BibitemOpen
  \bibfield  {author} {\bibinfo {author} {\bibfnamefont {A.~J.~K.}\
  \bibnamefont {Chua}},\ }\href {\doibase 10.1088/1742-6596/716/1/012028}
  {\bibfield  {journal} {\bibinfo  {journal} {J. Phys. Conf. Ser.}\ }\textbf
  {\bibinfo {volume} {716}},\ \bibinfo {pages} {012028} (\bibinfo {year}
  {2016})},\ \Eprint {http://arxiv.org/abs/1602.00620} {arXiv:1602.00620
  [gr-qc]} \BibitemShut {NoStop}%
\bibitem [{\citenamefont {Chua}\ \emph {et~al.}(2020)\citenamefont {Chua},
  \citenamefont {Korsakova}, \citenamefont {Moore}, \citenamefont {Gair},\ and\
  \citenamefont {Babak}}]{Chua:2019wgs}%
  \BibitemOpen
  \bibfield  {author} {\bibinfo {author} {\bibfnamefont {A.~J.~K.}\
  \bibnamefont {Chua}}, \bibinfo {author} {\bibfnamefont {N.}~\bibnamefont
  {Korsakova}}, \bibinfo {author} {\bibfnamefont {C.~J.}\ \bibnamefont
  {Moore}}, \bibinfo {author} {\bibfnamefont {J.~R.}\ \bibnamefont {Gair}}, \
  and\ \bibinfo {author} {\bibfnamefont {S.}~\bibnamefont {Babak}},\ }\href
  {\doibase 10.1103/PhysRevD.101.044027} {\bibfield  {journal} {\bibinfo
  {journal} {Phys. Rev. D}\ }\textbf {\bibinfo {volume} {101}},\ \bibinfo
  {pages} {044027} (\bibinfo {year} {2020})},\ \Eprint
  {http://arxiv.org/abs/1912.11543} {arXiv:1912.11543 [astro-ph.IM]}
  \BibitemShut {NoStop}%
\bibitem [{\citenamefont {Chua}\ and\ \citenamefont
  {Cutler}(2021)}]{chua2021non}%
  \BibitemOpen
  \bibfield  {author} {\bibinfo {author} {\bibfnamefont {A.~J.}\ \bibnamefont
  {Chua}}\ and\ \bibinfo {author} {\bibfnamefont {C.~J.}\ \bibnamefont
  {Cutler}},\ }\href@noop {} {\bibfield  {journal} {\bibinfo  {journal} {arXiv
  preprint arXiv:2109.14254}\ } (\bibinfo {year} {2021})}\BibitemShut {NoStop}%
\bibitem [{\citenamefont {Arnaud}\ \emph {et~al.}(2007)\citenamefont {Arnaud},
  \citenamefont {Auger}, \citenamefont {Babak}, \citenamefont {Baker},
  \citenamefont {Benacquista}, \citenamefont {Bloomer}, \citenamefont {Brown},
  \citenamefont {Camp}, \citenamefont {Cannizzo}, \citenamefont {Christensen},
  \citenamefont {Clark}, \citenamefont {Cornish}, \citenamefont {Crowder},
  \citenamefont {Cutler}, \citenamefont {Finn}, \citenamefont {Halloin},
  \citenamefont {Hayama}, \citenamefont {Hendry}, \citenamefont {Jeannin},
  \citenamefont {Kr{\'{o}}lak}, \citenamefont {Larson}, \citenamefont {Mandel},
  \citenamefont {Messenger}, \citenamefont {Meyer}, \citenamefont {Mohanty},
  \citenamefont {Nayak}, \citenamefont {Numata}, \citenamefont {Petiteau},
  \citenamefont {Pitkin}, \citenamefont {Plagnol}, \citenamefont {Porter},
  \citenamefont {Prix}, \citenamefont {Roever}, \citenamefont {Stroeer},
  \citenamefont {Thirumalainambi}, \citenamefont {Thompson}, \citenamefont
  {Toher}, \citenamefont {Umstaetter}, \citenamefont {Vallisneri},
  \citenamefont {Vecchio}, \citenamefont {Veitch}, \citenamefont {Vinet},
  \citenamefont {Whelan},\ and\ \citenamefont {Woan}}]{MLDC_1_Arnaud_2007}%
  \BibitemOpen
  \bibfield  {author} {\bibinfo {author} {\bibfnamefont {K.~A.}\ \bibnamefont
  {Arnaud}}, \bibinfo {author} {\bibfnamefont {G.}~\bibnamefont {Auger}},
  \bibinfo {author} {\bibfnamefont {S.}~\bibnamefont {Babak}}, \bibinfo
  {author} {\bibfnamefont {J.~G.}\ \bibnamefont {Baker}}, \bibinfo {author}
  {\bibfnamefont {M.~J.}\ \bibnamefont {Benacquista}}, \bibinfo {author}
  {\bibfnamefont {E.}~\bibnamefont {Bloomer}}, \bibinfo {author} {\bibfnamefont
  {D.~A.}\ \bibnamefont {Brown}}, \bibinfo {author} {\bibfnamefont {J.~B.}\
  \bibnamefont {Camp}}, \bibinfo {author} {\bibfnamefont {J.~K.}\ \bibnamefont
  {Cannizzo}}, \bibinfo {author} {\bibfnamefont {N.}~\bibnamefont
  {Christensen}}, \bibinfo {author} {\bibfnamefont {J.}~\bibnamefont {Clark}},
  \bibinfo {author} {\bibfnamefont {N.~J.}\ \bibnamefont {Cornish}}, \bibinfo
  {author} {\bibfnamefont {J.}~\bibnamefont {Crowder}}, \bibinfo {author}
  {\bibfnamefont {C.}~\bibnamefont {Cutler}}, \bibinfo {author} {\bibfnamefont
  {L.~S.}\ \bibnamefont {Finn}}, \bibinfo {author} {\bibfnamefont
  {H.}~\bibnamefont {Halloin}}, \bibinfo {author} {\bibfnamefont
  {K.}~\bibnamefont {Hayama}}, \bibinfo {author} {\bibfnamefont
  {M.}~\bibnamefont {Hendry}}, \bibinfo {author} {\bibfnamefont
  {O.}~\bibnamefont {Jeannin}}, \bibinfo {author} {\bibfnamefont
  {A.}~\bibnamefont {Kr{\'{o}}lak}}, \bibinfo {author} {\bibfnamefont {S.~L.}\
  \bibnamefont {Larson}}, \bibinfo {author} {\bibfnamefont {I.}~\bibnamefont
  {Mandel}}, \bibinfo {author} {\bibfnamefont {C.}~\bibnamefont {Messenger}},
  \bibinfo {author} {\bibfnamefont {R.}~\bibnamefont {Meyer}}, \bibinfo
  {author} {\bibfnamefont {S.}~\bibnamefont {Mohanty}}, \bibinfo {author}
  {\bibfnamefont {R.}~\bibnamefont {Nayak}}, \bibinfo {author} {\bibfnamefont
  {K.}~\bibnamefont {Numata}}, \bibinfo {author} {\bibfnamefont
  {A.}~\bibnamefont {Petiteau}}, \bibinfo {author} {\bibfnamefont
  {M.}~\bibnamefont {Pitkin}}, \bibinfo {author} {\bibfnamefont
  {E.}~\bibnamefont {Plagnol}}, \bibinfo {author} {\bibfnamefont {E.~K.}\
  \bibnamefont {Porter}}, \bibinfo {author} {\bibfnamefont {R.}~\bibnamefont
  {Prix}}, \bibinfo {author} {\bibfnamefont {C.}~\bibnamefont {Roever}},
  \bibinfo {author} {\bibfnamefont {A.}~\bibnamefont {Stroeer}}, \bibinfo
  {author} {\bibfnamefont {R.}~\bibnamefont {Thirumalainambi}}, \bibinfo
  {author} {\bibfnamefont {D.~E.}\ \bibnamefont {Thompson}}, \bibinfo {author}
  {\bibfnamefont {J.}~\bibnamefont {Toher}}, \bibinfo {author} {\bibfnamefont
  {R.}~\bibnamefont {Umstaetter}}, \bibinfo {author} {\bibfnamefont
  {M.}~\bibnamefont {Vallisneri}}, \bibinfo {author} {\bibfnamefont
  {A.}~\bibnamefont {Vecchio}}, \bibinfo {author} {\bibfnamefont
  {J.}~\bibnamefont {Veitch}}, \bibinfo {author} {\bibfnamefont {J.-Y.}\
  \bibnamefont {Vinet}}, \bibinfo {author} {\bibfnamefont {J.~T.}\ \bibnamefont
  {Whelan}}, \ and\ \bibinfo {author} {\bibfnamefont {G.}~\bibnamefont
  {Woan}},\ }\href {\doibase 10.1088/0264-9381/24/19/s16} {\bibfield  {journal}
  {\bibinfo  {journal} {Classical and Quantum Gravity}\ }\textbf {\bibinfo
  {volume} {24}},\ \bibinfo {pages} {S529} (\bibinfo {year}
  {2007})}\BibitemShut {NoStop}%
\bibitem [{\citenamefont {Babak}\ \emph {et~al.}(2008)\citenamefont {Babak}
  \emph {et~al.}}]{Babak:2008aa}%
  \BibitemOpen
  \bibfield  {author} {\bibinfo {author} {\bibfnamefont {S.}~\bibnamefont
  {Babak}} \emph {et~al.},\ }\href {\doibase 10.1088/0264-9381/25/18/184026}
  {\bibfield  {journal} {\bibinfo  {journal} {Class. Quant. Grav.}\ }\textbf
  {\bibinfo {volume} {25}},\ \bibinfo {pages} {184026} (\bibinfo {year}
  {2008})},\ \Eprint {http://arxiv.org/abs/0806.2110} {arXiv:0806.2110 [gr-qc]}
  \BibitemShut {NoStop}%
\bibitem [{\citenamefont {Babak}\ \emph {et~al.}(2010)\citenamefont {Babak}
  \emph {et~al.}}]{MockLISADataChallengeTaskForce:2009wir}%
  \BibitemOpen
  \bibfield  {author} {\bibinfo {author} {\bibfnamefont {S.}~\bibnamefont
  {Babak}} \emph {et~al.} (\bibinfo {collaboration} {Mock LISA Data Challenge
  Task Force}),\ }\href {\doibase 10.1088/0264-9381/27/8/084009} {\bibfield
  {journal} {\bibinfo  {journal} {Class. Quant. Grav.}\ }\textbf {\bibinfo
  {volume} {27}},\ \bibinfo {pages} {084009} (\bibinfo {year} {2010})},\
  \Eprint {http://arxiv.org/abs/0912.0548} {arXiv:0912.0548 [gr-qc]}
  \BibitemShut {NoStop}%
\bibitem [{\citenamefont {Amaro-Seoane}\ \emph {et~al.}(2011)\citenamefont
  {Amaro-Seoane}, \citenamefont {Schutz},\ and\ \citenamefont
  {Thornburg}}]{amaro2011gravitational}%
  \BibitemOpen
  \bibfield  {author} {\bibinfo {author} {\bibfnamefont {P.}~\bibnamefont
  {Amaro-Seoane}}, \bibinfo {author} {\bibfnamefont {B.}~\bibnamefont
  {Schutz}}, \ and\ \bibinfo {author} {\bibfnamefont {J.}~\bibnamefont
  {Thornburg}},\ }\href@noop {} {\bibfield  {journal} {\bibinfo  {journal}
  {arXiv preprint arXiv:1102.3647}\ } (\bibinfo {year} {2011})}\BibitemShut
  {NoStop}%
\bibitem [{\citenamefont {Chua}\ and\ \citenamefont
  {Gair}(2015)}]{AAK_Chua_2015_PRD}%
  \BibitemOpen
  \bibfield  {author} {\bibinfo {author} {\bibfnamefont {A.~J.~K.}\
  \bibnamefont {Chua}}\ and\ \bibinfo {author} {\bibfnamefont {J.~R.}\
  \bibnamefont {Gair}},\ }\href {\doibase 10.1088/0264-9381/32/23/232002}
  {\bibfield  {journal} {\bibinfo  {journal} {Classical and Quantum Gravity}\
  }\textbf {\bibinfo {volume} {32}},\ \bibinfo {pages} {232002} (\bibinfo
  {year} {2015})}\BibitemShut {NoStop}%
\bibitem [{\citenamefont {Osburn}\ \emph {et~al.}(2016)\citenamefont {Osburn},
  \citenamefont {Warburton},\ and\ \citenamefont {Evans}}]{osburn2016highly}%
  \BibitemOpen
  \bibfield  {author} {\bibinfo {author} {\bibfnamefont {T.}~\bibnamefont
  {Osburn}}, \bibinfo {author} {\bibfnamefont {N.}~\bibnamefont {Warburton}}, \
  and\ \bibinfo {author} {\bibfnamefont {C.~R.}\ \bibnamefont {Evans}},\
  }\href@noop {} {\bibfield  {journal} {\bibinfo  {journal} {Physical Review
  D}\ }\textbf {\bibinfo {volume} {93}},\ \bibinfo {pages} {064024} (\bibinfo
  {year} {2016})}\BibitemShut {NoStop}%
\bibitem [{\citenamefont {Van De~Meent}(2018)}]{van2018gravitational}%
  \BibitemOpen
  \bibfield  {author} {\bibinfo {author} {\bibfnamefont {M.}~\bibnamefont {Van
  De~Meent}},\ }\href@noop {} {\bibfield  {journal} {\bibinfo  {journal}
  {Physical Review D}\ }\textbf {\bibinfo {volume} {97}},\ \bibinfo {pages}
  {104033} (\bibinfo {year} {2018})}\BibitemShut {NoStop}%
\bibitem [{\citenamefont {Hughes}\ \emph {et~al.}(2021)\citenamefont {Hughes},
  \citenamefont {Warburton}, \citenamefont {Khanna}, \citenamefont {Chua},\
  and\ \citenamefont {Katz}}]{hughes2021adiabatic}%
  \BibitemOpen
  \bibfield  {author} {\bibinfo {author} {\bibfnamefont {S.~A.}\ \bibnamefont
  {Hughes}}, \bibinfo {author} {\bibfnamefont {N.}~\bibnamefont {Warburton}},
  \bibinfo {author} {\bibfnamefont {G.}~\bibnamefont {Khanna}}, \bibinfo
  {author} {\bibfnamefont {A.~J.}\ \bibnamefont {Chua}}, \ and\ \bibinfo
  {author} {\bibfnamefont {M.~L.}\ \bibnamefont {Katz}},\ }\href@noop {}
  {\bibfield  {journal} {\bibinfo  {journal} {Physical Review D}\ }\textbf
  {\bibinfo {volume} {103}},\ \bibinfo {pages} {104014} (\bibinfo {year}
  {2021})}\BibitemShut {NoStop}%
\bibitem [{\citenamefont {Chua}\ \emph {et~al.}(2019)\citenamefont {Chua},
  \citenamefont {Galley},\ and\ \citenamefont {Vallisneri}}]{chua2019reduced}%
  \BibitemOpen
  \bibfield  {author} {\bibinfo {author} {\bibfnamefont {A.~J.}\ \bibnamefont
  {Chua}}, \bibinfo {author} {\bibfnamefont {C.~R.}\ \bibnamefont {Galley}}, \
  and\ \bibinfo {author} {\bibfnamefont {M.}~\bibnamefont {Vallisneri}},\
  }\href@noop {} {\bibfield  {journal} {\bibinfo  {journal} {Physical review
  letters}\ }\textbf {\bibinfo {volume} {122}},\ \bibinfo {pages} {211101}
  (\bibinfo {year} {2019})}\BibitemShut {NoStop}%
\bibitem [{\citenamefont {Luo}\ \emph {et~al.}(2016{\natexlab{a}})\citenamefont
  {Luo}, \citenamefont {Chen}, \citenamefont {Duan}, \citenamefont {Gong},
  \citenamefont {Hu}, \citenamefont {Ji}, \citenamefont {Liu}, \citenamefont
  {Mei}, \citenamefont {Milyukov}, \citenamefont {Sazhin} \emph
  {et~al.}}]{luo2016tianqin}%
  \BibitemOpen
  \bibfield  {author} {\bibinfo {author} {\bibfnamefont {J.}~\bibnamefont
  {Luo}}, \bibinfo {author} {\bibfnamefont {L.-S.}\ \bibnamefont {Chen}},
  \bibinfo {author} {\bibfnamefont {H.-Z.}\ \bibnamefont {Duan}}, \bibinfo
  {author} {\bibfnamefont {Y.-G.}\ \bibnamefont {Gong}}, \bibinfo {author}
  {\bibfnamefont {S.}~\bibnamefont {Hu}}, \bibinfo {author} {\bibfnamefont
  {J.}~\bibnamefont {Ji}}, \bibinfo {author} {\bibfnamefont {Q.}~\bibnamefont
  {Liu}}, \bibinfo {author} {\bibfnamefont {J.}~\bibnamefont {Mei}}, \bibinfo
  {author} {\bibfnamefont {V.}~\bibnamefont {Milyukov}}, \bibinfo {author}
  {\bibfnamefont {M.}~\bibnamefont {Sazhin}},  \emph {et~al.},\ }\href@noop {}
  {\bibfield  {journal} {\bibinfo  {journal} {Classical and Quantum Gravity}\
  }\textbf {\bibinfo {volume} {33}},\ \bibinfo {pages} {035010} (\bibinfo
  {year} {2016}{\natexlab{a}})}\BibitemShut {NoStop}%
\bibitem [{\citenamefont {Hu}\ \emph {et~al.}(2018)\citenamefont {Hu},
  \citenamefont {Li}, \citenamefont {Wang}, \citenamefont {Feng}, \citenamefont
  {Zhou}, \citenamefont {Hu}, \citenamefont {Hu}, \citenamefont {Mei},\ and\
  \citenamefont {Shao}}]{Hu2018}%
  \BibitemOpen
  \bibfield  {author} {\bibinfo {author} {\bibfnamefont {X.-C.}\ \bibnamefont
  {Hu}}, \bibinfo {author} {\bibfnamefont {X.-H.}\ \bibnamefont {Li}}, \bibinfo
  {author} {\bibfnamefont {Y.}~\bibnamefont {Wang}}, \bibinfo {author}
  {\bibfnamefont {W.-F.}\ \bibnamefont {Feng}}, \bibinfo {author}
  {\bibfnamefont {M.-Y.}\ \bibnamefont {Zhou}}, \bibinfo {author}
  {\bibfnamefont {Y.-M.}\ \bibnamefont {Hu}}, \bibinfo {author} {\bibfnamefont
  {S.-C.}\ \bibnamefont {Hu}}, \bibinfo {author} {\bibfnamefont {J.-W.}\
  \bibnamefont {Mei}}, \ and\ \bibinfo {author} {\bibfnamefont {C.-G.}\
  \bibnamefont {Shao}},\ }\href {\doibase 10.1088/1361-6382/aab52f} {\bibfield
  {journal} {\bibinfo  {journal} {Class. Quant. Grav.}\ }\textbf {\bibinfo
  {volume} {35}},\ \bibinfo {pages} {095008} (\bibinfo {year} {2018})},\
  \Eprint {http://arxiv.org/abs/1803.03368} {arXiv:1803.03368 [gr-qc]}
  \BibitemShut {NoStop}%
\bibitem [{\citenamefont {Luo}\ \emph {et~al.}(2016{\natexlab{b}})\citenamefont
  {Luo} \emph {et~al.}}]{Luo2015}%
  \BibitemOpen
  \bibfield  {author} {\bibinfo {author} {\bibfnamefont {J.}~\bibnamefont
  {Luo}} \emph {et~al.} (\bibinfo {collaboration} {TianQin}),\ }\href {\doibase
  10.1088/0264-9381/33/3/035010} {\bibfield  {journal} {\bibinfo  {journal}
  {Class. Quant. Grav.}\ }\textbf {\bibinfo {volume} {33}},\ \bibinfo {pages}
  {035010} (\bibinfo {year} {2016}{\natexlab{b}})},\ \Eprint
  {http://arxiv.org/abs/1512.02076} {arXiv:1512.02076 [astro-ph.IM]}
  \BibitemShut {NoStop}%
\bibitem [{\citenamefont {Liu}\ \emph {et~al.}(2020)\citenamefont {Liu},
  \citenamefont {Hu}, \citenamefont {Zhang}, \citenamefont {Mei} \emph
  {et~al.}}]{liu2020science}%
  \BibitemOpen
  \bibfield  {author} {\bibinfo {author} {\bibfnamefont {S.}~\bibnamefont
  {Liu}}, \bibinfo {author} {\bibfnamefont {Y.-M.}\ \bibnamefont {Hu}},
  \bibinfo {author} {\bibfnamefont {J.-d.}\ \bibnamefont {Zhang}}, \bibinfo
  {author} {\bibfnamefont {J.}~\bibnamefont {Mei}},  \emph {et~al.},\
  }\href@noop {} {\bibfield  {journal} {\bibinfo  {journal} {Physical Review
  D}\ }\textbf {\bibinfo {volume} {101}},\ \bibinfo {pages} {103027} (\bibinfo
  {year} {2020})}\BibitemShut {NoStop}%
\bibitem [{\citenamefont {Babak}(2008)}]{babak2008building}%
  \BibitemOpen
  \bibfield  {author} {\bibinfo {author} {\bibfnamefont {S.}~\bibnamefont
  {Babak}},\ }\href@noop {} {\bibfield  {journal} {\bibinfo  {journal}
  {Classical and Quantum Gravity}\ }\textbf {\bibinfo {volume} {25}},\ \bibinfo
  {pages} {195011} (\bibinfo {year} {2008})}\BibitemShut {NoStop}%
\bibitem [{\citenamefont {Babak}\ \emph
  {et~al.}(2009{\natexlab{b}})\citenamefont {Babak}, \citenamefont {Gair},\
  and\ \citenamefont {Porter}}]{babak2009}%
  \BibitemOpen
  \bibfield  {author} {\bibinfo {author} {\bibfnamefont {S.}~\bibnamefont
  {Babak}}, \bibinfo {author} {\bibfnamefont {J.~R.}\ \bibnamefont {Gair}}, \
  and\ \bibinfo {author} {\bibfnamefont {E.~K.}\ \bibnamefont {Porter}},\
  }\href@noop {} {\bibfield  {journal} {\bibinfo  {journal} {Classical and
  quantum gravity}\ }\textbf {\bibinfo {volume} {26}},\ \bibinfo {pages}
  {135004} (\bibinfo {year} {2009}{\natexlab{b}})}\BibitemShut {NoStop}%
\bibitem [{\citenamefont {Wang}\ \emph
  {et~al.}(2012{\natexlab{b}})\citenamefont {Wang}, \citenamefont {Shang},\
  and\ \citenamefont {Babak}}]{wang2012extreme}%
  \BibitemOpen
  \bibfield  {author} {\bibinfo {author} {\bibfnamefont {Y.}~\bibnamefont
  {Wang}}, \bibinfo {author} {\bibfnamefont {Y.}~\bibnamefont {Shang}}, \ and\
  \bibinfo {author} {\bibfnamefont {S.}~\bibnamefont {Babak}},\ }\href@noop {}
  {\bibfield  {journal} {\bibinfo  {journal} {Physical Review D}\ }\textbf
  {\bibinfo {volume} {86}},\ \bibinfo {pages} {104050} (\bibinfo {year}
  {2012}{\natexlab{b}})}\BibitemShut {NoStop}%
\bibitem [{\citenamefont {Riles}(2023)}]{riles2023searches}%
  \BibitemOpen
  \bibfield  {author} {\bibinfo {author} {\bibfnamefont {K.}~\bibnamefont
  {Riles}},\ }\href@noop {} {\bibfield  {journal} {\bibinfo  {journal} {Living
  Reviews in Relativity}\ }\textbf {\bibinfo {volume} {26}},\ \bibinfo {pages}
  {3} (\bibinfo {year} {2023})}\BibitemShut {NoStop}%
\bibitem [{\citenamefont {Prix}\ and\ \citenamefont
  {Shaltev}(2012)}]{prix2012search}%
  \BibitemOpen
  \bibfield  {author} {\bibinfo {author} {\bibfnamefont {R.}~\bibnamefont
  {Prix}}\ and\ \bibinfo {author} {\bibfnamefont {M.}~\bibnamefont {Shaltev}},\
  }\href@noop {} {\bibfield  {journal} {\bibinfo  {journal} {Physical Review
  D}\ }\textbf {\bibinfo {volume} {85}},\ \bibinfo {pages} {084010} (\bibinfo
  {year} {2012})}\BibitemShut {NoStop}%
\bibitem [{\citenamefont {Skilling}(2004)}]{skilling2004nested}%
  \BibitemOpen
  \bibfield  {author} {\bibinfo {author} {\bibfnamefont {J.}~\bibnamefont
  {Skilling}},\ }in\ \href@noop {} {\emph {\bibinfo {booktitle} {Aip conference
  proceedings}}},\ Vol.\ \bibinfo {volume} {735}\ (\bibinfo {organization}
  {American Institute of Physics},\ \bibinfo {year} {2004})\ pp.\ \bibinfo
  {pages} {395--405}\BibitemShut {NoStop}%
\bibitem [{\citenamefont {Speagle}(2020)}]{speagle2020dynesty}%
  \BibitemOpen
  \bibfield  {author} {\bibinfo {author} {\bibfnamefont {J.~S.}\ \bibnamefont
  {Speagle}},\ }\href@noop {} {\bibfield  {journal} {\bibinfo  {journal}
  {Monthly Notices of the Royal Astronomical Society}\ }\textbf {\bibinfo
  {volume} {493}},\ \bibinfo {pages} {3132} (\bibinfo {year}
  {2020})}\BibitemShut {NoStop}%
\bibitem [{\citenamefont {Speri}\ \emph {et~al.}(2023)\citenamefont {Speri},
  \citenamefont {Katz}, \citenamefont {Chua}, \citenamefont {Hughes},
  \citenamefont {Warburton}, \citenamefont {Thompson}, \citenamefont
  {Chapman-Bird},\ and\ \citenamefont {Gair}}]{Speri2023FastAF}%
  \BibitemOpen
  \bibfield  {author} {\bibinfo {author} {\bibfnamefont {L.}~\bibnamefont
  {Speri}}, \bibinfo {author} {\bibfnamefont {M.~L.}\ \bibnamefont {Katz}},
  \bibinfo {author} {\bibfnamefont {A.~J.}\ \bibnamefont {Chua}}, \bibinfo
  {author} {\bibfnamefont {S.~A.}\ \bibnamefont {Hughes}}, \bibinfo {author}
  {\bibfnamefont {N.}~\bibnamefont {Warburton}}, \bibinfo {author}
  {\bibfnamefont {J.~E.}\ \bibnamefont {Thompson}}, \bibinfo {author}
  {\bibfnamefont {C.~E.~A.}\ \bibnamefont {Chapman-Bird}}, \ and\ \bibinfo
  {author} {\bibfnamefont {J.~R.}\ \bibnamefont {Gair}},\ }\href
  {https://api.semanticscholar.org/CorpusID:260125021} {\bibfield  {journal}
  {\bibinfo  {journal} {Frontiers Appl. Math. Stat.}\ }\textbf {\bibinfo
  {volume} {9}} (\bibinfo {year} {2023})}\BibitemShut {NoStop}%
\bibitem [{\citenamefont {Littenberg}\ and\ \citenamefont
  {Cornish}(2023)}]{littenberg2023prototype}%
  \BibitemOpen
  \bibfield  {author} {\bibinfo {author} {\bibfnamefont {T.~B.}\ \bibnamefont
  {Littenberg}}\ and\ \bibinfo {author} {\bibfnamefont {N.~J.}\ \bibnamefont
  {Cornish}},\ }\href@noop {} {\bibfield  {journal} {\bibinfo  {journal}
  {Physical Review D}\ }\textbf {\bibinfo {volume} {107}},\ \bibinfo {pages}
  {063004} (\bibinfo {year} {2023})}\BibitemShut {NoStop}%
\bibitem [{\citenamefont {Cornish}\ \emph {et~al.}(2021)\citenamefont
  {Cornish}, \citenamefont {Littenberg}, \citenamefont {B{\'e}csy},
  \citenamefont {Chatziioannou}, \citenamefont {Clark}, \citenamefont
  {Ghonge},\ and\ \citenamefont {Millhouse}}]{cornish2021bayeswave}%
  \BibitemOpen
  \bibfield  {author} {\bibinfo {author} {\bibfnamefont {N.~J.}\ \bibnamefont
  {Cornish}}, \bibinfo {author} {\bibfnamefont {T.~B.}\ \bibnamefont
  {Littenberg}}, \bibinfo {author} {\bibfnamefont {B.}~\bibnamefont
  {B{\'e}csy}}, \bibinfo {author} {\bibfnamefont {K.}~\bibnamefont
  {Chatziioannou}}, \bibinfo {author} {\bibfnamefont {J.~A.}\ \bibnamefont
  {Clark}}, \bibinfo {author} {\bibfnamefont {S.}~\bibnamefont {Ghonge}}, \
  and\ \bibinfo {author} {\bibfnamefont {M.}~\bibnamefont {Millhouse}},\
  }\href@noop {} {\bibfield  {journal} {\bibinfo  {journal} {Physical Review
  D}\ }\textbf {\bibinfo {volume} {103}},\ \bibinfo {pages} {044006} (\bibinfo
  {year} {2021})}\BibitemShut {NoStop}%
\bibitem [{\citenamefont {van~der Walt}\ \emph {et~al.}(2011)\citenamefont
  {van~der Walt}, \citenamefont {Colbert},\ and\ \citenamefont
  {Varoquaux}}]{vanderWalt:2011bqk}%
  \BibitemOpen
  \bibfield  {author} {\bibinfo {author} {\bibfnamefont {S.}~\bibnamefont
  {van~der Walt}}, \bibinfo {author} {\bibfnamefont {S.~C.}\ \bibnamefont
  {Colbert}}, \ and\ \bibinfo {author} {\bibfnamefont {G.}~\bibnamefont
  {Varoquaux}},\ }\href {\doibase 10.1109/MCSE.2011.37} {\bibfield  {journal}
  {\bibinfo  {journal} {Comput. Sci. Eng.}\ }\textbf {\bibinfo {volume} {13}},\
  \bibinfo {pages} {22} (\bibinfo {year} {2011})},\ \Eprint
  {http://arxiv.org/abs/1102.1523} {arXiv:1102.1523 [cs.MS]} \BibitemShut
  {NoStop}%
\bibitem [{\citenamefont {Virtanen}\ \emph {et~al.}(2020)\citenamefont
  {Virtanen} \emph {et~al.}}]{Virtanen:2019joe}%
  \BibitemOpen
  \bibfield  {author} {\bibinfo {author} {\bibfnamefont {P.}~\bibnamefont
  {Virtanen}} \emph {et~al.},\ }\href {\doibase 10.1038/s41592-019-0686-2}
  {\bibfield  {journal} {\bibinfo  {journal} {Nature Meth.}\ }\textbf {\bibinfo
  {volume} {17}},\ \bibinfo {pages} {261} (\bibinfo {year} {2020})},\ \Eprint
  {http://arxiv.org/abs/1907.10121} {arXiv:1907.10121 [cs.MS]} \BibitemShut
  {NoStop}%
\bibitem [{\citenamefont {Hunter}(2007)}]{Hunter:2007ouj}%
  \BibitemOpen
  \bibfield  {author} {\bibinfo {author} {\bibfnamefont {J.~D.}\ \bibnamefont
  {Hunter}},\ }\href {\doibase 10.1109/MCSE.2007.55} {\bibfield  {journal}
  {\bibinfo  {journal} {Comput. Sci. Eng.}\ }\textbf {\bibinfo {volume} {9}},\
  \bibinfo {pages} {90} (\bibinfo {year} {2007})}\BibitemShut {NoStop}%
\end{thebibliography}%

\end{document}